\documentclass[12pt]{article}

\usepackage[round]{natbib}
\usepackage{amsmath}
\usepackage{actuarialsymbol}
\usepackage{url}
\usepackage{graphicx}
\usepackage{caption}
\usepackage{subcaption}
\usepackage[a4paper, margin=2.5cm]{geometry}
\usepackage[pacote de cores]{xcolor}
\usepackage{lscape}
\usepackage{array}
\usepackage{multirow}
\usepackage[toc,page]{appendix}
\usepackage[utf8]{inputenc}
\usepackage{authblk}
\usepackage{ulem}
\usepackage{bm}
\usepackage{amsfonts}

\def\E{{\mathbb E}}

\def\Var{\mathbb{VAR}}

\title{How life-table right-censoring affected the Brazilian Social Security Factor: an application of the gamma-Gompertz-Makeham model}
\author[1]{Filipe Costa de Souza\thanks{filipe.costas@ufpe.br}}
\author[1]{Wilton Bernardino}
\author[2]{Silvio C. Patricio}
\affil[1]{\small{Universidade Federal de Pernambuco}}
\affil[2]{\small{The Interdisciplinary Centre on Population Dynamics, University of Southern Denmark}}

\date{}

\begin{document}
	\maketitle

\textbf{Abstract}: Automatic Adjustment Mechanisms (AAM) are legal instruments that help social security systems respond to demographic and economic changes. In Brazil, the Social Security Factor (SSF) was introduced in the late 1990s as an AAM to link retirement benefits to life expectancy at the retirement age, with the hope of promoting contributory justice and discouraging early retirement. Recent research has highlighted the limitations of right-censored life tables, such as those used in Brazil. It has recommended using the gamma-Gompertz-Makeham ($\Gamma GM$) model to estimate adult and old-age mortality. This study investigated the impact of right-censoring on the SSF by comparing the official SSF and other social security metrics with a counterfactual scenario computed based on fitted $\Gamma GM$ models. The results indicate that from 2004 to 2012, official life tables may have negatively impacted retirees' income, particularly for those who delayed their retirement. Furthermore, the $\Gamma GM$ fitted models' life expectancies had more stable paths over time, which could have helped with long-term planning. This study's findings are significant for policymakers as they highlight the importance of using appropriate mortality metrics in AAMs to ensure accurate retirement benefit payments. They also underscore the need to consider the potential impacts of seemingly innocuous hypotheses on public action outcomes. Overall, this study provides valuable insights for public planners and policymakers looking to enhance the effectiveness and fairness of social security systems.


 
\section{Introduction}
	
Mortality measures extracted from national life tables, such as life expectancy at a given age, play a critical role in public planning and policy-making \citep{Lledo2017}, especially within the context of social security. The linkage between life expectancy and pension benefits or statutory retirement age has become increasingly prevalent throughout the globe, with automatic adjustment mechanisms (AAMs) being implemented in response to longevity improvements \citep{Tavernier2021, Ayuso2021}. In Brazil, the Law n. 9,876/1999 introduced the so-called Brazilian Social Security Factor (SSF) (\textit{fator previdenciário}), an AAM that linked pension benefit to life expectancy at the retirement age \citep{Lima2012, Lourenco2017}, with the hope to promote contributory justice and to discourage early retirements \citep{Cechin2007}.
	
In the recent past, a worker claiming for retirement by time of contribution in the Brazilian General Social Security Scheme (\textit{Regime Geral de Previdência Social - RGPS}) would have his or her retirement benefit computed based on a formula that includes the SSF. In its formulation, the SSF had three variables, namely: (i) the time of contribution to the system, (ii) the retirement age and (iii) the life expectancy at that retirement age. The latter should be computed based on the last (period and both-sex combined) complete life table published by the Brazilian Institute of Geography and Statistics (\textit{Instituto Brasileiro de Geografia e Estatística - IBGE}). 
	
The credibility of mortality measures, such as life expectancy, computed based on right-censured life tables was questioned by \cite{Missov2016} and \cite {Nemeth2018}, mainly when the open-ended age interval contains a large proportion of survivors. According to the authors, it is a frequent practice in countries with low-quality mortality data to aggregate life-table information after an arbitrary age (e.g., 80 years and above) and to suppose a constant-hazard hypothesis in the open-ended age interval when computing the life expectancy at the last available age \citep{Preston2001}, as a closing procedure to the life table. \cite{Missov2016} also adverted that when the last available age in the life table is too low, the constant-hazard hypothesis in the open-ended age interval becomes unrealistic and probably leads to distorted (high) life expectancies. In this respect, the authors argued their discussion based on Brazilian data.
	
The consequences of right-censoring due to low-quality data at old ages have been discussed since the 1980s. \cite{Horiuchi1982} developed a method to deal with life expectancy overestimation. The process assumes that above the last available age, the mortality is described by a Gompertz model and that the annual rate of increase of the population at the open-ended interval is known. This approach was improved by \cite{Mitra1984} when information about the mean age of the population in the open-ended interval was available. 

Despite the relevance of the contributions from \cite{Horiuchi1982} and \cite{Mitra1984}, the most popular alternative to deal with right-censoring is the extrapolation approach \citep{Ediev2018a, Ediev2018b}, which consists of fitting young and adult mortality data to a parametric mortality law and then extrapolating the death rates to old ages based on the fitted model. This approach has some advantages over \cite{Horiuchi1982} and \cite{Mitra1984} methods: (i) the parameters of the mortality model have demographic meaning \citep{Nemeth2018}, (ii) the model could provide age-specific death rates for ages above the last available age \citep{Ediev2018b}, (iii) Mortality metrics, such as life expectancy, could be computed for non-integer ages \citep{Nemeth2018}; and (iv) mortality measures calculated from parametric models are less sensitive to the censoring age \citep{Missov2016}.       
	
Regarding the choice of the parametric model, \cite{Missov2015} argue in favor of the gamma-Gompertz-Makeham ($\Gamma GM$) model, which is flexible in capturing the increasing risk of death at the adult ages and the leveling-off of the risk of dying at a level of 0.7  \citep{alvarez2021regularities, Barbi2018, gampe2010human, gampe2021mortality}. The $\Gamma GM$ model is a more flexible version of the Kannisto model \citep{Thatcher1998}, used by the Human Mortality Database, that allows for any positive asymptote, and provides a good fit for the force of mortality at the oldest ages \citep{missov2023improvements}. 
		
Due to the importance that life expectancy has in social security planning and inspired by the studies of \cite{Missov2016} and \cite{Nemeth2018}, this paper aims to investigate the impact that right-censoring had in the SSF from 2000 to 2019, comparing for each year the actual SSF and other social security metrics, such as the normal retirement age, computed based on life expectancies provided by IBGE life tables and those of counterfactual scenarios where life expectancies were computed based on the parameters of fitted $\Gamma GM$ models.

This paper contributes to the Brazilian social security debate, adding new insights about the SSF and the impact that an apparently harmless assumption could have on the government and retirees. Moreover, the investigation provides some insights into using parametric models (particularly the $\Gamma GM$ model) to estimate old-age mortality in Brazil. Therefore, we understand that this study contributes to the literature on (i) the impact that implicit assumptions in life table construction can have on life expectancy \citep{Missov2016, Lledo2017, Nemeth2018} and how it could reverberate to social security results and planning \citep{Wilbert2013}, (ii) the literature about public pension reforms in Brazil \citep{Beltrao2014, Caetano2016, Afonso2019, Lima2019} and the importance of adopting AAM in this process \citep{Vidal2009, Ayuso2021b, Tavernier2021} (iii) and the literature about the use of parametric models for old-age mortality estimation in Brazil \citep{Franco2007, Gonzaga2022}.
	
The remaining sections of the paper are organized as follows: Section \ref{sec:SSF} describes the Brazilian SSF formulation and historical background, as well other social security rules that coexisted with or replaced it (\ref{sec:SSF_rule}), and also presents previous discussions and criticisms in the literature about the SSF (\ref{sec:studies}). Section \ref{sec:MM} describes the main elements of a life table (\ref{sec:life table}), the data used in the investigation (\ref{sec:data}), the $\Gamma GM$ model and the estimation strategy (\ref{sec:GGM}) and the main analyses performed in the paper (\ref{sec:analyses}). Section \ref{sec:Results} presents the paper's main results. In Section \ref{sec:Discussion}, those findings are discussed in the light of the literature. Finally, Section \ref{sec:Conclusion} provides the investigation's conclusions.  
	
\section{The Brazilian Social Security Factor}\label{sec:SSF}

\subsection{The SSF rule}\label{sec:SSF_rule}
	
According to \cite{Ribeiro2008}, the Brazilian pension system presented a positive balance at the time of promulgation of the 1988 Federal Constitution. Nevertheless, this situation has changed since the mid-1990s, and growing deficits started to be observed, which caused discussions about the necessity of reforms in the Brazilian Social Security System. Retirement by time of contribution was at the center of those debates since it allowed retirement at young ages with high pension benefits \citep{Caetano2016}.
	
In 1995, a Constitutional Amendment Proposal was sent to the Brazilian Congress, having as one of its main goals, originally, to establish a minimal retirement age rule.	However, after three years of discussion, the proposal of a minimal retirement age rule was not approved \citep{Caetano2016}. In December 1998, the Constitutional Amendment n. 20 was approved, establishing, among other things, that the social security system should be financially and actuarially balanced and the deconstitutionalization of the rule for computing retirement benefits, which opened space for the promulgation of the law n. 9,876/1999 that established the SSF \citep{Lima2012, Lourenco2017}. 
	
Overall, bounded by some legal limits, the SSF aimed at linking pension benefits to contributions (promoting contributory justice) and discouraging early retirements \citep{Cechin2007}, hoping to help balance the system's accounts. Thus, the SSF is an AAM \citep{Vidal2009, Tavernier2021} that links pension benefits to period life expectancy at the retirement age. In the RGPS, the application of the SSF was mandatory for retirement by the time of contribution and optional for retirement by age. By the Law n. 9,876/1999 \citep{lei9876}, the SSF for a person retiring at age $x$ was defined as 
\begin{equation}\label{factor}
	\textrm{SSF}_{x, CT} = \frac{CT\times A}{e_{x}}\left(1+\frac{x+CT \times A}{100}\right), 
\end{equation}	
where $CT$ represents the contribution time at the moment of retirement, $A$ is a constant value equal to 0.31 representing the contribution rate (11\% from the employee and 20\% from the employer\footnote{In practice, overall, the employer contributes with 20\% over the employee's actual salary, even if the salary of the latter is higher than the system's ceiling. On the other hand, the employees contribute with a rate defined according to some contribution salary ranges, i.e., that must respect the system's minimum and maximum benefits, and could be equal to 8\%, 9\% or 11\% \citep{Penafieri2013}}), $x$ is the insured's age at the time of retirement, $e_{x}$ is the period (both-sex combined) life expectancy at the time of retirement computed based on the last available life table published by IBGE \citep{Brasil1999}.
	
In the RGPS, the minimum effective contribution time for men was 35 years, while for women, it was 30 years. However, in the calculation of the SSF for women, there was a bonus $\beta=5$ years, which should be added to the effective contribution time ($ECT$) to the system,  i.e.,  when computing the SSF, it should be assumed that $CT=ECT+\beta$. Moreover, male teachers could retire with 30 years of contribution (receiving a bonus of 5 years), and female teachers could retire with 25 years of contribution (receiving a bonus of 10 years).
	
The pension benefit for retirement by the time of contribution was defined as 
\begin{equation}\label{benefit}
	B_{x, CT} = \max\left[\min\left(\textrm{SSF}_{x, CT} \times M, C\right), W\right], 
\end{equation}	
where $M$ defines the (monetarily adjusted) mean of the 80\% highest salaries of contribution computed between July 1994 and the moment of retirement\footnote{In December 2022, the Brazilian Supreme Court (\textit{Supremo Tribunal Federal}), after a long judicial dispute, recognized the right for those that entered at the social security system before the Law n. 9,876/1999 of using all the contributory period (i.e., including periods before July 1994), if financially advantageous.}, $C$ denotes the ceiling (maximal possible value) for pension benefits in the RGPS, and $W$ is the minimum wage. Additionally, from December 1999 to November 2004 (the transition period), before being fully applied, the SSF went through 60 months of a transition rule \citep{Lima2012}, with the transition factor being computed as 
\begin{equation}\label{transition} 
	f_{n} = \frac{\textrm{SSF}_{x, CT}\times n}{60}+\frac{60-n}{60},
\end{equation}
where $n$ denotes the number of months between the promulgation of the law n. 9,879/1999 and the retirement moment, and $f_{n}$ was the transition factor. Moreover, during the transition period, the calculation of benefits for retirement by time of contribution was defined according to Equation~\eqref{benefit}, by replacing $\textrm{SSF}_{x, CT}$ to $f_{n}$.
	
\cite{Pinheiro1999} explain the context in which the SSF rule was conceived, which helps the understanding of its formula. The SSF was inspired by the non-financial (or notional) defined contribution (NDC) schemes implemented in some European countries such as Sweden, Italy and Latvia in the 1990s \citep{Holzmann2017}. However, according to \cite{Pinheiro1999}, to implement a NDC scheme in Brazil the government would need to overcome two main problems: (i) the political discussion about the system's rate of return and the monetary correction rates, and (ii) operational aspects concerning administrative data.
	
According to the authors, the real interest rate of 6\% per year, which was usually adopted as a standard real rate of return for actuarial calculations by the complementary pension funds in Brazil at that time, would be used as a baseline level in Brazilian Congress discussions, which would lead to serious sustainability problems to the system. Moreover, the high inflation in the 1980s and early 1990s would create difficulties in establishing the proper monetary correction rates, leaving the system vulnerable to judicial actions. Finally, the low reliability or lack of information for earlier periods would make the reconstruction of contributions accumulated by the workers very difficult. \citep{Pinheiro1999}.
	
The rate of return issue was handled by making the accumulated contributions by workers to be capitalized based on a rate that considers the contribution time to the system and the retirement age. The two other issues were dealt with by establishing that the retirement benefit would be calculated based on the mean of the 80\% highest salaries of contribution computed between July 1994 and the moment of retirement, a period in which the administrative data would have better quality and in which inflation would have been controlled by the Real Plan \citep{Pinheiro1999}. Thus, the SSF formula could be interpreted as follows: in the first part of the formula, the term $CT \times A$ indicates the total contribution made in favor of the worker, and the division of this term by the $e_{x}$ equalizes the contributions to the expected time in retirement. The second part of the formula works as an interest rate that remunerates the insured for more time of contribution to the pension system, or for delaying the retirement moment \citep{Pinheiro1999, Cechin2007, Ribeiro2008}.
	
The SSF has faced criticism since its inception, resulting in political and social pressure to establish new retirement rules that can either replace or at least serve as an alternative for the SSF rule. Despite previous attempts \citep{Lima2012, Penafieri2013}, it was only in June 2015, with the Provisional Measure n. 676 and its conversion, in November 2015, into the law n. 13,183, that a new retirement rule began to coexist with the SSF, and it was called the 85/95 progressive rule. By this new retirement rule, the worker whose sum of age and effective contribution time was greater than or equal to 85 (for females) or 95 (for males) would retire without the application of the SSF, i.e., once the conditions were met, the SSF would only be applied if it benefited the retiree. Moreover, male and female teachers received a bonus of 5 years to be added to the sum of age and contribution time\footnote{It is also important to remember that the minimum effective contribution times were 35 years for men, 30 years for women and male teachers and 25 years for female teachers.}. The 85/95 progressive rule also had a discretionary adjustment mechanism, in which the values 85 and 95 should gradually progress every two years until they reached the limit of 90 and 100, respectively, in December 2026 \citep{Caetano2016, Lima2019}.

\cite{Lima2012} pointed that the SSF was unable to reverse the trend of higher increasing of expenditures in relation to the revenues and, therefore, the Brazilian pension system would remain unbalanced. However, \cite{Caetano2016} argue that although SSF was not able to balance the system account, it showed itself as an important tool to slow down the rate at which the RGPS expenses increased, and that the promulgation of the 85/95 rule distanced the Brazilian social security system from the best international sustainability practices. This view is also shared by \cite{Lima2019}. \cite{Afonso2019} also emphasize that the 85/95 rule was a populist measure and that it went against the efforts to balance the system. This fact accentuated the debate on deeper reforms in the Brazilian pension system, especially about the definition of a minimum retirement age rule, which would return to the political debate in Temer and Bolsonaro governments.
	
In 2016, Michel Temer's government sent to the Congress the Constitutional Amendment Proposal n. 287, which ended up not being voted due to the military intervention in Rio de Janeiro, and due to the proximity of the presidential elections. In 2019, after the election of Jair Bolsonaro, a new social security reform was approved throughout the Constitutional Amendment n. 103. This reform established a minimum retirement age rule putting an end to the retirement by time of contribution and, consequently, to the SSF (and to the short-lived 85/95 rule). Nevertheless, there are two cases in which the SSF can still be applied. The first case is for the insureds who, at the time of the reform, had already completed the necessary requirements to retire by the previous rules (SSF or 85/95) and, therefore, could retain the right to retire under those rules. In this case, for those opting to retire under the SSF rule, the SSF to be used will be the one of the year 2019, the year of the reform. The other case in which the SSF can be applied is a transition rule for those who have already completed 28 years of contribution (for females) or 33 years (for males) at the time of the reform. Those workers will only be able to retire when they achieve 30 years of contribution (for females) and 35 years (for males) plus an additional period of 50\% of the time left to achieve the 30 years of contribution (for females) and 35 years (for males). In this case, the retirement benefit can be computed based on the average of all contribution salaries, between July 1994 and the moment of retirement, multiplied by the SSF \citep{Brasil2019}.

\subsection{Related studies}\label{sec:studies}

As stated, ever since its creation in 1999, the SSF attracted political and social criticism and was the focus of academic research, and the use of life expectancy extracted from IBGE life tables was a critical part of this debate. Overall, investigations discuss the theoretical background of the SSF formula \citep{Cechin2007, Penafieri2013}, economic and legal reflex that could be caused by the use of different life tables or methodological updates \citep{Deud2004, Meneguin2015, Wilbert2013}, the fiscal sustainability and distributive aspects of the pension system \citep{Caetano2006, Caetano2016, Lima2012, Lima2019, Freitas2019}, among other issues.  

According to \cite{Cechin2007}, although the SSF has promoted an advance when compared to the previous retirement rules in Brazil, its formula still carried some issues. For example, the authors argued that the SSF left risk benefits unfunded, as it would use the entire contribution rate to fund the planned retirement. Additionally, they also pointed that the contribution rate adopted as a standard in its formula was higher than the average contribution rate of the system, and the life expectancies used in the calculation of the SSF should be those that best describe the survival pattern of the system's insureds, and not those describing the whole Brazilian population. In what concerns this last observation, \cite{Wilbert2013} investigated the impact that different life tables could have on RGPS pension benefits estimation. Their study compared the 2002 IBGE life table with a life table built based on administrative data of RGPS retirees (which had higher life expectancies), and concluded that the difference in life expectancy between the two life tables represented a negative impact on the RGPS deficit of approximately 4.5\%. Moreover, \cite{Ribeiro2008} also point that the SSF ignores the mortality heterogeneity of Brazilian population and, therefore, it may perpetuate some social inequalities. 

\cite{Penafieri2013} argue that even though the SSF had a clear goal, it lacked a theoretical actuarial background, a view that was also shared by \cite{Beltrao2014}. \cite{Penafieri2013} showed that the SSF reduced (increased) early (late) pension benefits more than the actuarial fair level, which should not be viewed as a problem due to the main goal of the mechanism. \cite{Delgado2006} and \cite{Caetano2006} indicated that the SSF rule was able to increase the average retirement age and to reduce the average value of the pension benefit for those retiring by time of contribution, since it discouraged early retirement, providing some economy to the RGPS expenses. \cite{Caetano2006} also highlighted that the SSF combined with other legal mechanisms, such as contributions rates and eligibility conditions, made social security system progressive, also acting as a compensatory tool for inequalities in the labor market. However, \cite{Lima2012} observed that the SSF was not able to balance the pension system since the expenses were still growing faster than the revenues, also emphasizing that the SSF was only applied as a mandatory rule for the retirement by time of contribution.     

Another important criticism about the SSF came from the work of \cite{Deud2004}, who highlighted that sudden changes in the methodology for the construction of Brazilian life tables could cause legal insecurity for workers planning their retirement and even lead to legal disputes. The author recommended that demographic research centers in Brazil should monitor and revise the construction of IBGE life tables, providing greater transparency and credibility to the results. According to \cite{Meneguin2015}, the Brazilian National Institute for Social Security (\textit{Instituto Nacional de Seguro Social - INSS}) is the largest litigant in the entire Brazilian Judiciary, and according to \cite{Baars2012} the SSF was the main cause of re-retirement (\textit{desaposentação}) \citep{Zanella2014} legal claims.

Even though essencial investigations and discussions about the SSF and the use of IBGE life tables have been proposed, the literature still lacks investigations that use parametric models to discuss the impact of life-tables closing procedures in the calculation of life expectancies and, consequently, on the SSF (which could also impact on other dimensions of public planning). Additionally, understanding some limitations of the SSF can help in the planning of new AAM and prevent mistakes from being repeated, primarily because past experiences would most likely be remembered in new political debates on pension reform, and future legislation could also continue to adopt IBGE tables as a standard.
	
\section{Material and Methods}\label{sec:MM}
	
\subsection{Life tables}\label{sec:life table}
	
Overall, a period life table is an instrument that summarizes, in a given period of time, for a sequence of ages, the distribution of deaths of a synthetic cohort \citep{Pitacco2009}. Traditionally, for a sequence of integer and non-negative ages, a period life table provides the following elements: the number of persons alive at the exact age $x$, $l_{x}$; the number of deaths between exact ages $x$ and $x+n$, $_{n}d_{x} = l_{x}-l_{x+n}$; the probability of a person age $x$ survives to at least age $x+n$, $\px[n]{x} = \frac{l_{x+n}}{l_{x}}$; the probability of a person age $x$ dies before age $x+n$, $_{n}q_{x} = 1-\px[n]{x}$; the average person-years lived by those dying between exact ages $x$ and $x+n$, $_{n}a_{x}$; the number of person-years lived between exact ages $x$ and $x+n$, $_{n}L_{x} = n \times l_{x+n}+{_{n}a_{x}} \times {_{n}d_{x}}$; the age-specific death rate between exact ages $x$ and $x+n$, $_{n}m_{x} = \frac{_{n}d_{x}}{_{n}L_{x}}$; the person-years lived above age $x$, $T_{x} = \sum_{j=x}^{\infty}{_{n}L_{j}}$; and the period life expectancy at exact age $x$, $e_{x} = \frac{T_{x}}{l_{x}}$ \citep{Preston2001}. Moreover, if $n=1$, i.e. the life table is complete, we can simply omit the underwritten symbol on the left side, e.g., $_{1}p_{x}$ is simply written as $p_{x}$.     
	
While constructing a complete life table, it is conventionally assumed that there is a final age $\omega$, in which no one is expected to live beyond, i.e., $l_{x} = 0, \forall x\geq \omega$. Another alternative is to aggregate the information about deaths up to an arbitrary age, $\overline{x}$ (or $x+$), providing complete life tables that are right-censored, i.e., that have an open-ended age interval. As stated by \cite{Preston2001}, a standard assumption to close a life table in an open-ended age interval is to assume that ${_\infty}a_{\overline{x}}=e_{\overline{x}}=\frac{1}{{_\infty}m_{\overline{x}}}$, where $_{\infty}d_{\overline{x}}=l_{\overline{x}}$ and $_{\infty}L_{\overline{x}}=T_{\overline{x}}={_\infty}a_{\overline{x}}.{_\infty}d_{\overline{x}}$. This means that, for all ages in the open-ended interval, persons are exposed to a constant hazard of death \citep{Missov2016}.
	
%
%
	
\subsection{Data}\label{sec:data}
	
According to the Decree 3,266/1999 \citep{Brasil1999} and for purposes of calculating the SSF \citep{lei9876}, it was incumbent upon IBGE to publish, by December 1st, the Brazilian life tables of the previous year. Thus, annually and since 1999, IBGE publishes Brazilian official complete life tables. Since the Brazilian official life tables for a given calendar year $t$ are only published at the end of the subsequent year, i.e., at the end of calendar year $t+1$, overall, the SSF computed in a given calendar year $t+2$ uses life expectancies gathered from the life table of calendar year $t$. For example, overall, the 2012 SSF was computed using the 2010 IBGE life table. 
	
For the investigation, complete period (both-sex combined) life tables from 1998 to 2017 were gathered freely from IBGE official website (\url{ibge.gov.br}). IBGE complete life tables are built from abridged ones, and mortality information are aggregated at the open-ended interval of 80 years and beyond. However, for some years, IBGE life tables were published with some inconsistencies that needed to be revised for a proper use in our investigation. For example, in 2006 $d_{80} = 41,982$ while $l_{80}=41,849$ and for some ages $d_{x}$ was different from $l_{x} - l_{x+1}$ (e.g. $d_{79}=2,675$ and $l_{79}=44,649$). Therefore, before fitting the $\Gamma GM$ models, we reconstructed the $d_{x}$ and $L_{x}$ functions from $l_{x}$ \citep{Shkolnikov2017}. Moreover, for all ages up to 79, $L_{x}$ was calculated based on the standard uniform distribution of deaths (UDD) assumption \citep{Dickson2019}.  
	
During the period under investigation, the life tables published by IBGE underwent two major changes and, therefore, can be categorized into three periods of analysis: 1998-2001, 2002-2010 and 2011 onward. The four published life tables for the calendar years 1998 to 2001 were projected tables computed based on the 1991 Brazilian life table (i.e., that used information from the 1991 census) and a limit life table adopted by the U.S. Bureau of the Census \citep{IBGE2002}. The life tables from 2002 to 2010 used information from the 2000 population census and were built based on the components method \citep{IBGE2004}. Finally, the 2011 and further Brazilian life tables were constructed supported by data from the 2010 demographic census. Moreover, the abridged life tables started to adopt the 90+ open-ended interval, instead the 80+ open-ended interval adopted by the previous abridged tables \citep{IBGE2013}. Further methodological details, including the methods to adjust under-registration of deaths and procedures to obtain a complete life table from a abridged one, can be found at \cite{IBGE2002, IBGE2004, IBGE2013, IBGE2016}.     
	
\begin{figure}[h]
	\centering
	\includegraphics[scale=0.8]{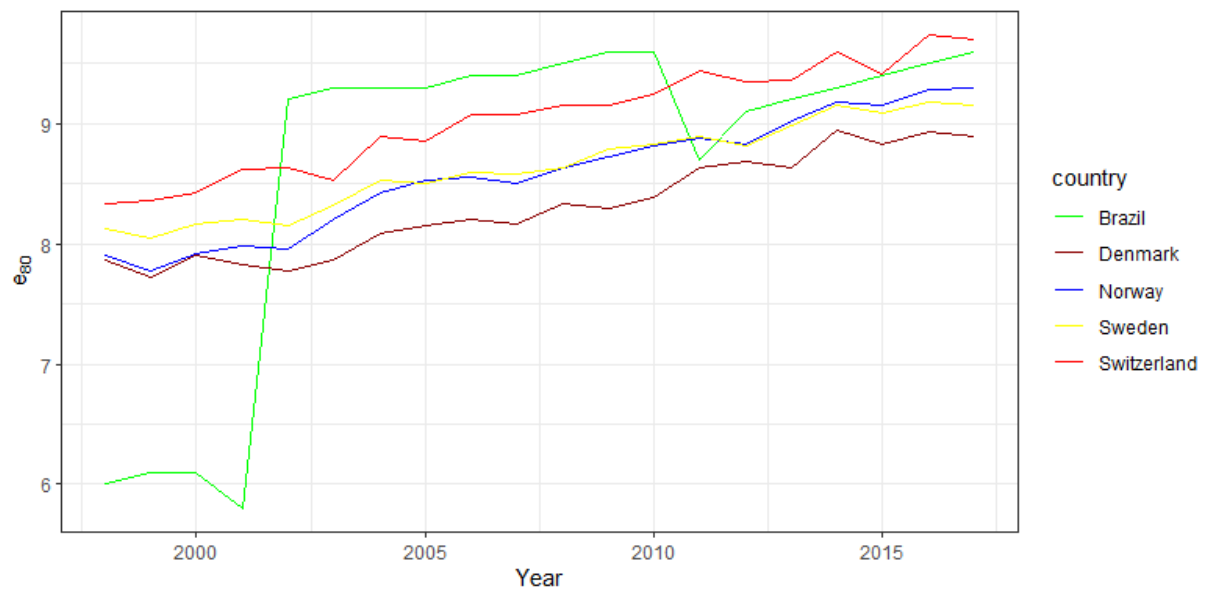}
	\caption{Life expectancy at 80 years, a comparison between Brazil, Denmark, Norway, Sweden and Switzerland, 1998-2017. Data source: Brazil: www.ibge.gov.br; Denmark,  Norway, Sweden and Switzerland: mortality.org.}
	\label{fig:e80}
\end{figure}

For the sake of empirical motivation, Figure \ref{fig:e80} compares the life expectancies at 80 years between Brazil and some selected countries known for the longevity of their populations and for the high quality of their mortality datasets. The Brazilian data were gathered from IBGE complete life tables, while the remaining data were gathered form the Human Mortality Database (HMD) (\url{mortality.org}), which smooths old age mortality based on the Kannisto model \citep{Wilmoth2021, Thatcher1998}. From 2002 to 2010, it can be seen that the $e_{80}$ taken from IBGE tables were higher than those of Danish, Norwegian, Swedish and Swiss people, and from 2012 to 2017 they were only slightly lower than those from Switzerland. Based on the 2003 life tables, for example, $e_{80}$ in Brazil was approximately 1.4, 1.1, 1.0 and 0.8 years higher than the ones of the populations mentioned above, respectively. This evidence reinforces the argument that Brazilian life expectancies were overestimated, at least since 2002 (and especially from 2002 to 2010). Also, it allows us to observe how the age adopted to close a life table influences life expectancies since, as stated, until 2010 Brazilian abridged tables were closed at 80+ and in 2011 they began to be closed at 90+.   
	
%

\subsection{The gamma-Gompertz-Makeham model}\label{sec:GGM}

We use the gamma-Gompertz-Makeham ($\Gamma GM$) model to smooth the mortality rates. The model incorporates three key components: the Gompertz function, which describes the age-dependent increase in mortality risk; the Makeham parameter, which captures the age-independent mortality risk, such as genetic predisposition to disease or exposure to environmental hazards; and the gamma distribution, which models the individual-level heterogeneity in mortality risk (i.e., frailty) that is not explained by age or other observable factors \citep{Vaupel1979}. 

The force of mortality at the age $x$ for the $\Gamma GM$ model is given by:
\begin{equation} \label{eq:ggomp}
	\mu(x| \bm \theta) = \frac{ae^{bx}}{1+\sigma^2 \frac{a}{b}\left( e^{bx}-1\right)} + c
\end{equation}
where $a > 0$ represents the initial level of mortality risk at age zero; $b > 0$ represents the rate of increase in mortality risk with age and determines the shape of the age-dependent mortality curve; $c \geq 0$ is the Makeham parameter that captures the age-independent mortality risk and represents the baseline mortality rate that is not dependent on age; and $\sigma^2 > 0$ represents the shape of the frailty distribution and determines the degree of heterogeneity in mortality risk, capturing the individual-level heterogeneity in mortality risk that is not explained by age or other observable factors. 

Some additional notation is in order. Let $D_x$ be the number of deaths in  a given age interval $[x,x+1)$ for $x=0,\ldots,m$. Also, let $E_x$ denote the number of person-years with age $x$ exposed to the risk of dying \citep[see, for example,][]{brillinger1986biometrics, macdonald2018modelling, castellares2022gompertz}. Also, define $\bm{D}=(D_0, D_1, \dots, D_m)^\top$ and $\bm{E}=(E_0, E_1, \dots, E_m)^\top$. In addition, let $\bm{\theta}=(a,b,c, \sigma^2)^\top$ be the parameter vector that characterizes the force of mortality at age $x$ which is given by Equation~\eqref{eq:ggomp}. Finally, we assume that the number of deaths and the number of person-years exposed to the risk of dying can be observed.

The standard approach to estimate mortality models was presented by \cite{brillinger1986biometrics}. It assumes a Poisson distribution for the number of deaths $D_x$ with $\E(D_x) = \Var(D_x) = \mu(x| \bm \theta) E_x$. Therefore, the probability mass function of the random variable $D_x$ is 
\begin{equation*}
	\mathbb{P}[D_x=z]=\frac{e^{-\mu(x| \bm \theta) E_x}\,(\mu(x| \bm \theta) E_x)^{z}}{z!}, z\in \{0,1,2,\dots\}.
\end{equation*}

By considering the standard assumption on the death count, the likelihood function for the parameter vector $\bm{\theta}=(a,b,c, \sigma^2)^\top$ is given by
\begin{equation*}
	L(\bm{\theta}) \equiv L(\bm{\theta}|\bm{D},\bm{E}) =
	\prod_{x}\frac{e^{-\mu(x| \bm \theta) E_x}\,(\mu(x| \bm \theta) E_x)^{D_x}}{D_x!},
\end{equation*}
and the log-likelihood function takes the form
\begin{equation*}
	\ell(\bm{\theta})\equiv\ell(\bm{\theta}|\bm{D},\bm{E})=
	\ln(L(\bm{\theta})) = \sum_{x}[D_x\ln(\mu(x| \bm \theta) E_x)
	-\mu(x| \bm \theta) E_x].
\end{equation*}
The maximum likelihood (ML) estimator $\hat{\bm{\theta}} = \left(\hat{a}, \hat{b}, \hat{c}, \hat{\sigma^2} \right) ^\top$ of $\bm \theta = (a, b, c, \sigma^2)^\top$ is obtained by maximizing the log-likelihood function with respect to $\bm \theta = (a, b, c, \sigma^2)^\top$.

There is no closed-form expression for the ML estimator $\widehat{\bm{\theta}}$; therefore, its computation must be performed numerically using a nonlinear optimization algorithm. The maximization of the log-likelihood function can be performed, for example, using the {\sf R} programming language \citep{R2023}, specifically applying the {\tt optim} function or the Differential Evolution Optimization algorithm \citep{mullen2011deoptim}. In our analysis, we used mortality after age 30 and the ML estimator $\widehat{\bm{\theta}}$ to calculate the remaining life expectancies at age $x$ through closed-form expression provided by \cite{Castellares2020}. 
	
\subsection{Analyses}\label{sec:analyses}
	
Before explaining the primary proposed analyses, it is essential to clarify some technical aspects concerning the use of life expectancy that influence the calculation of the SSF, which may not be completely clear in the legal text. When computing the SSF, the contribution time $CT$ and the retirement age $x$ could assume non-integer values. However, once the official life tables only provide life expectancies for integer ages, then the SSF formula uses the life expectancy at the age $\lfloor x \rfloor$, where $\lfloor . \rfloor$ indicates the floor function. Moreover, life expectancies at the retirement ages gathered from the official life tables are rounded to one decimal digit to be consistent with the life expectancies published every year by the IBGE at the Union Official Gazette (\textit{Diário Oficial da União}). In our investigation, this same approach was adopted when computing a SSF. To provide an example of the SSF calculation, Tables \ref{tab:SSF_IBGE} and \ref{tab:SSF_GGM} (in the Appendices) present a fragment of the 2012 SSF tables built from the 2010 IBGE life table and the 2010 fitted $ \Gamma GM$ model, respectively.  

The analyses were carried out for every year between 2000 and 2019, and the calculations were performed on  {\sf R}. Moreover, since the 85/95 progressive rule started to coexist with the SSF rule in the mid 2015, then, for that calendar year the analyses were made assuming both situations: the SSF rule alone and the coexistence of both rules. Furthermore, in 2019 the 85 and 95 values were adjusted to 86 and 96 respectively.     
	
Both from the perspective of the retirees and the government, a central point of discussion would be how life expectancy computed by a fitted $\Gamma GM$ model (the counterfactual scenario) could affect the SSF and consequently the pension benefit when compared to official SSF computed from IBGE life tables. This inquiry holds significance as benefit improvements (or reductions) may influence workers' retirement decision and wealth, as well as public spending and revenue, and could also generate distortions in the labor market \citep{Queiroz2021, Souza2019}. Therefore, it was calculated the relative discrepancy (\%) between the SSF computed by the official life table ($\textrm{SSF}_{x, CT}^{IBGE}$) and by the fitted $\Gamma GM$ model ($\textrm{SSF}_{x, CT}^{\Gamma GM}$),
	
\begin{equation}\label{prop}
	i_{x, CT}=\left( \frac{\textrm{SSF}_{x, CT}^{\Gamma GM}}{\textrm{SSF}_{x, CT}^{IBGE}}-1\right) \times 100.
\end{equation}
	
Before the 85/95 progressive rule was introduced in mid-2015 (or if we are considering solely the SSF rule), Equation~\eqref{prop} would only depend on age. This means that it could be simply written as  $\left(\frac{e_{\lfloor x \rfloor}^{IBGE}}{e_{\lfloor x \rfloor}^{\Gamma GM}}-1 \right) \times 100$, i.e., it would be the relative discrepancy between life expectancies. Additionally, between mid-2015 to 2019, retirees who met the criteria to retire under the 85/95 progressive rule, the SSF would be that maximum value between 1 and the result from Equation~\eqref{factor}, and, consequently, Equation~\eqref{prop} also depends on $CT$. Furthermore, it is important to highlight that if $i_{x, CT}$ is equal to 5\% for a given scenario, it does not mean that a retiree would receive a pension benefit 5\% higher if the $\Gamma GM$ model were used instead of the official tables. As pointed out in Equation~\eqref{benefit}, the retirement benefit has legal limits and, therefore, the value $i_{x, CT}$ should be interpreted as the maximum relative discrepancy. Finally, comparisons were not carried out on the transition factors, Equation~\eqref{transition}, which also depends on how many months have passed after the enactment of the Law n. 9,876/1999 and, consequently, could create additional complications that would deviate the focus from the central analysis.    
	
Another essential social security metric concerns to the Normal Retirement Age (NRA) \citep{Tavernier2021}. Assuming a person that entered the labor market at a given age $y$, the NRA is the age at which this citizen could, after a full career (i.e. the effective contribution time is given by $ECT=NRA-y$) and considering any eventual bonus ($\beta$), retire with full pension benefit, i.e. with SSF equal to 1. The NRAs were computed assuming 18 and 23 as the starting ages. So, for each type of worker\footnote{Male worker, female worker, male teacher and female teacher.} and respecting the legal constraints, it is necessary to find the age $x$ that solves Equation~\eqref{NRA}, 
	
\begin{equation}\label{NRA}
	SSF_{x}= \frac{(x-y+\beta) \times A}{e_{\lfloor x \rfloor}} \times \left(1 + \frac{x+(x-y+\beta) \times A}{100}\right) =1.
\end{equation}
	
Similarly, for each type of worker and respecting the legal constraints, it could also be questioned, for each integer age $x$, what is the contribution time that makes the SSF equals to 1 ($CT1$)? Thus, for each age $x$, it would be necessary to find the $CT$ that solves Equation~\eqref{CT1},
	
\begin{equation}\label{CT1}
	SSF_{x, CT}= \frac{CT \times A}{e_{\lfloor x \rfloor}} \times \left(1 + \frac{x+CT \times A}{100}\right) = 1.
\end{equation}
	
\section{Results}\label{sec:Results}

\begin{figure}[h]
	\centering
	\includegraphics[scale=0.7]{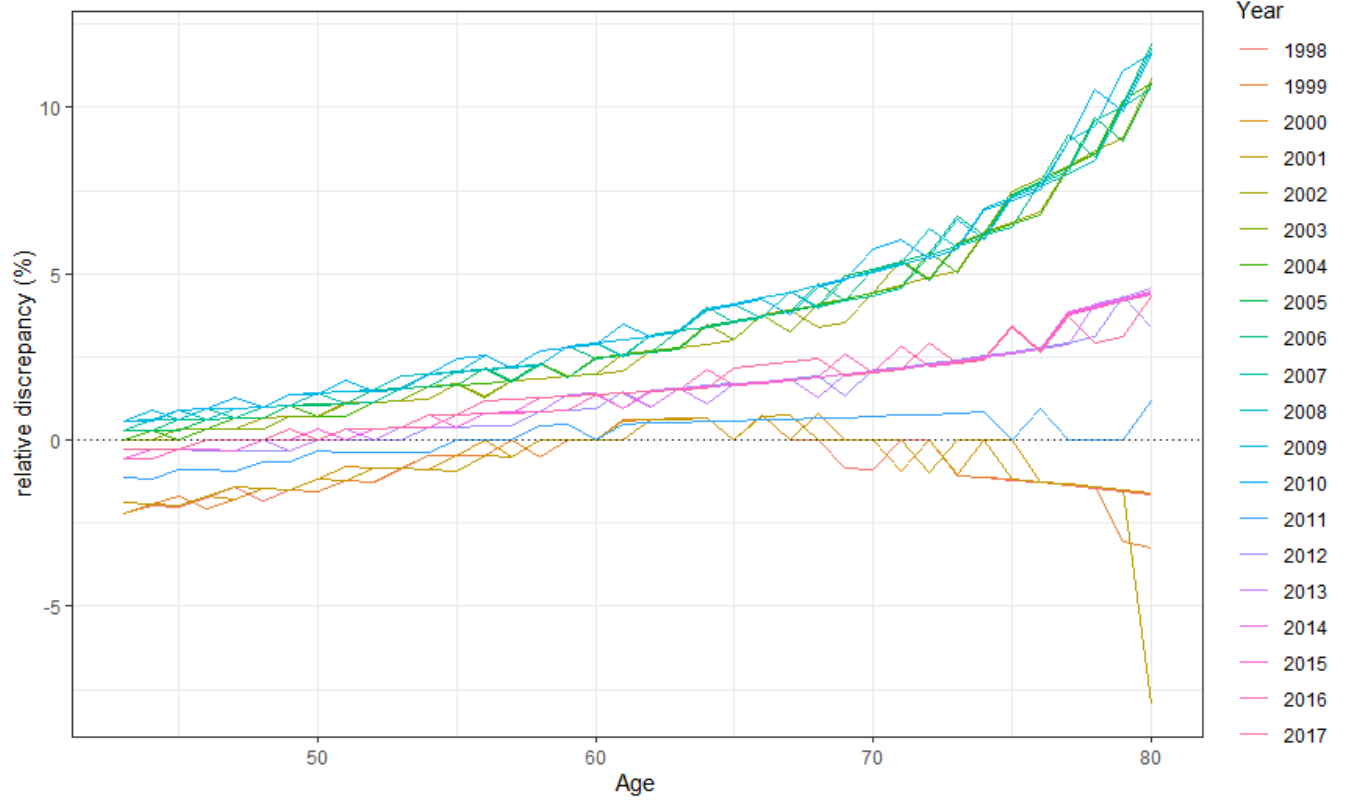}
	\caption{The relative discrepancy between life expectancies from the IBGE life tables and $\Gamma GM$ fitted models, for ages from 43 to 80.}
	\label{fig:relative_discrepancy}
\end{figure}

The first step to compare the SSF computed from the IBGE tables and the $\Gamma GM$ fitted models is to analyze the relative discrepancy between life expectancies produced by these two scenarios for the possible integer retirement ages (i.e., $x\geq 43$ years), during the period of analysis. Figure \ref{fig:relative_discrepancy} displays the relative discrepancy between life expectancies computed from IBGE life table and those estimated based on $\Gamma GM$ models, from 1998 to 2017. This initial analysis can also be extended to describe the relative discrepancy between the official and the counterfactual SSF, as discussed in Section \ref{sec:analyses}. 

According to data from 1998 to 2001, for all ages between 43 and 60 years and between 69 and 80 years, life expectancies gathered from the IBGE tables were lower than or equal to the life expectancies predicted by the fitted $\Gamma GM$ models. Between ages 61 to 68 life expectancies obtained from IBGE tables were greater than or equal to those calculated from the $\Gamma GM$ fitted models. Furthermore, during this period, while considering public planning and the need to discourage early retirement, it is worth noting that as life expectancies presented in IBGE tables appear to be lower than they really would be \citep{Oliveira2003}. In this regard, the $\Gamma GM$ model would be advantageous as it shows life expectancies greater than or equal to those of the official life tables for all ages from 43 to 60 years. However, the effects on pension benefits would be reduced once during this period the transition factor was being applied.

From the 2002 to the 2010 life tables, the scenario was quite different, especially because official life expectancies for retirement ages appear to be overestimated according to more recent official estimates  \citep{IBGE2013b, IBGE2018}. In this period, for all ages between 43 and 80 years, life expectancies from official tables were greater than or equal to those from the fitted $\Gamma GM$ models, and the relative discrepancy increased with age, reaching values greater than 10\% at the age of 80 years. This fact implies that, from 2004 to 2012, a citizen who was retiring by the time of contribution would be in a better situation (or at least as good as) if the SSF was computed using the life expectancy from a fitted $ \Gamma GM$ model than the life expectancies of the respective official life table. From the 2004 life table to the 2010 life table (used to compute the SSF from 2006 to 2012), for a person retiring at 65 years, for example, the relative discrepancy between the SSFs was at least 3.5\%, reaching values greater than 4\% from 2010 to 2012, which could represent a sizable increasing on the retirement income. 

\begin{figure}[h]
	\centering
	\includegraphics[scale=0.7]{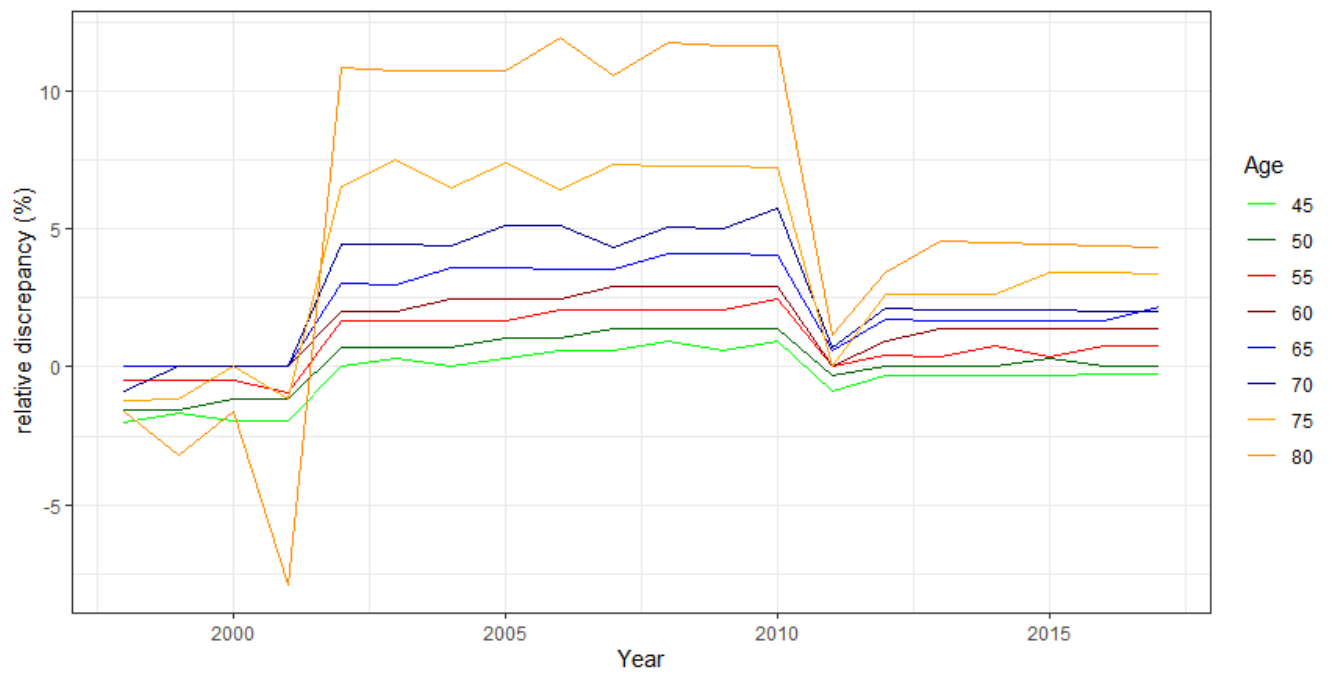}
	\caption{The yearly relative discrepancy between life expectancies from the IBGE life tables and from the $\Gamma GM$ models, from 1998 to 2017, for the ages of $45, 50, \dots, 80$.}
	\label{fig:relative_discrepancy_ages}
\end{figure}

The chart in Figure \ref{fig:relative_discrepancy_ages} describes the yearly relative discrepancy between the life expectancy calculated from IBGE life tables and those from the $\Gamma GM$ fitted models, for some selected ages ($45, 50, \dots, 80$), analyzing data from 1998 to 2017. Moreover,  despite the methodological change implemented in 2002, the transition between life expectancies from 2001 to 2002 was more seamless with $\Gamma GM$ fitted models than the official life tables. This could have helped to alleviated political and social pressures encountered that time, as stated by \cite{Deud2004}. 

Figure \ref{fig:relative_discrepancy_ages} helps to understand the following highlights. From the 2011 life tables, life expectancies from official life tables (built from abridged ones closed at the open-ended interval 90+) and the $\Gamma GM$ fitted models started to present lower differences when compared with those of the previous period. It was observed that in 2011, during the initial year of the new methodology, life expectancies calculated using $\Gamma GM$ modeling were higher than those obtained from official life tables for ages ranging from 43 to 54 years. Between the ages of 55 to 80 years, life expectancies estimated by the $\Gamma GM$ method were either lower or equal to those provided by IBGE\footnote{It is also important to highlight that life expectancies from ages 55 to 80 were higher in the 2010 official table than those from the 2011 table, fact that did not occur with the fitted models. In this particular case, even if life expectancies reductions could be considered positive from retiree's point of view, it brings instability and planning insecurity both for workers and government and helps to highlight the limitation of the traditional method of closing mortality tables.}. From the 2012 to the 2015 tables, life expectancies gathered from the official tables were lower than or equal to those gathered from the $\Gamma GM$ fitted models for ages in the range 43-49, and they were greater than or equal to those from the fitted models from 50 years onward. Between 2016 and 2017, official life expectancies for individuals aged 43 to 45 were either lower or equal to those computed using the $\Gamma GM$ fitted models, however, they were either higher or equal to those computed using the $\Gamma GM$ fitted models for people aged 49 and above. It is worth mentioning that from 2011 to 2017, the $\Gamma GM$ models predicted longer life expectancies for younger people. Nevertheless, the $\Gamma GM$ estimates showed lower life expectancy predictions for older people. This fact would have helped, at least theoretically, to discourage early and reward late retirements. 

\begin{figure}[h]
	\centering
	\includegraphics[scale=0.85]{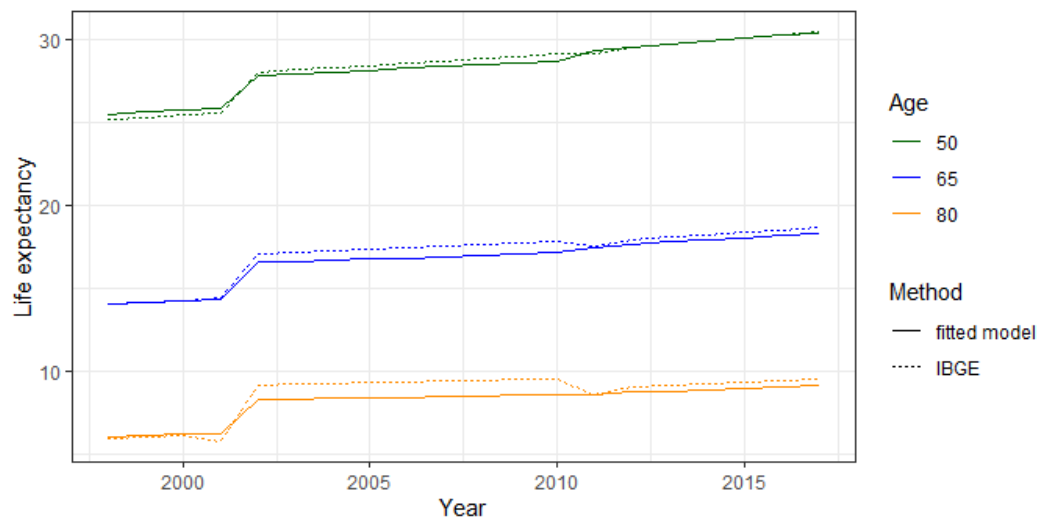}
	\caption{Comparison between life expectancies at ages 50, 65 and 80, from IBGE life tables and the $\Gamma GM$ fitted models, from 1998 to 2017.}
	\label{fig:ex_c}
\end{figure}

Figure \ref{fig:ex_c} shows the life expectancies at ages 50, 65 and 80, from 1998 to 2017, under the official and the counterfactual scenarios. This allow us to observe how those life expectancies evolve over time in each scenario, confirming that the $\Gamma GM$ models resulted in more stable paths. It is also worth noticing that in 2002, the life expectancy at 80 years computed from IBGE life table was equal to 9.2 years. This same life expectancy was only reached by the $\Gamma GM$ fitted models in 2017, a 15-year delay. Additionally, the life expectancy at 65 years old was 17.1, a value that the $\Gamma GM$ fitted models only reached seven years later.

It is also worth emphasizing that starting in mid-2015, the SSF rule began to coexist with the 85/95 progressive rule. This means that there would be no difference in retirement benefits (even if there were some differences in life expectancies) for those who retired under the 85/95 progressive rule in either scenarios. Moreover, the only differences in the value of the retirement benefits could exist for cases in which the worker retired under the SSF rule in both scenarios, or if it was better to him or her to retire under the SSF rule in one scenario and under the 85/95 progressive rule in the other. In the first case, the maximum difference between the retirement benefits would be given by the relative discrepancy between the life expectancies, as has been discussed before. In addition, for each calendar year, it can be observed that there were only a few specific combinations of age and $ECT$ in which the 85/95 progressive rule was a better option under the official life table scenario, but it would be better to retire under the SSF rule if the $\Gamma GM$ fitted model was adopted. For example, consider a female worker who was 63 years old and had 31 years of effective contribution. Thus, based on the 2013 official life table, her 2015 SSF would be 0.992 under the SSF rule and 1 under the 85/95 progressive rule, so she should retire under the second rule (see Table \ref{tab:SSF_IBGE_85} in the Appendix). However, considering the $\Gamma GM$ fitted model, her 2015 SSF would be 1.007 under the SSF rule and 1 under the 85/95 progressive rule, thus she would prefer to retire under the first rule (see Table \ref{tab:SSF_GGM_85} in the Appendix). This would only result in a 0.7\% increase in the pension benefit when comparing both scenarios, as shown in Table \ref{tab:SSF_GGM} in the Appendix.   

\begin{figure}[h]
	\centering
	\begin{subfigure}{0.48\textwidth}
		\centering
		\includegraphics[width=\textwidth]{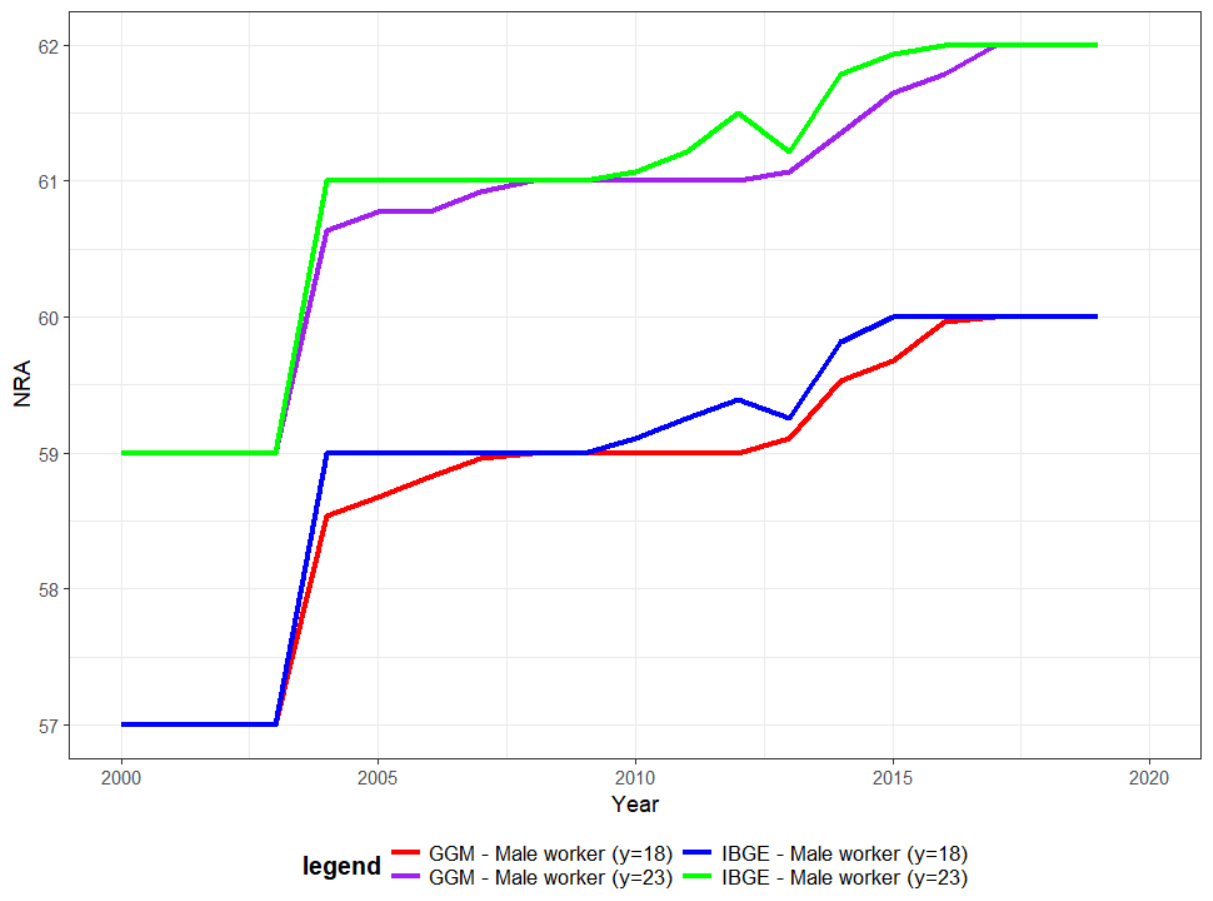}
		\caption{}\label{fig5a}
	\end{subfigure}
	\hfill
	\begin{subfigure}{0.48\textwidth}
		\centering
		\includegraphics[width=\textwidth]{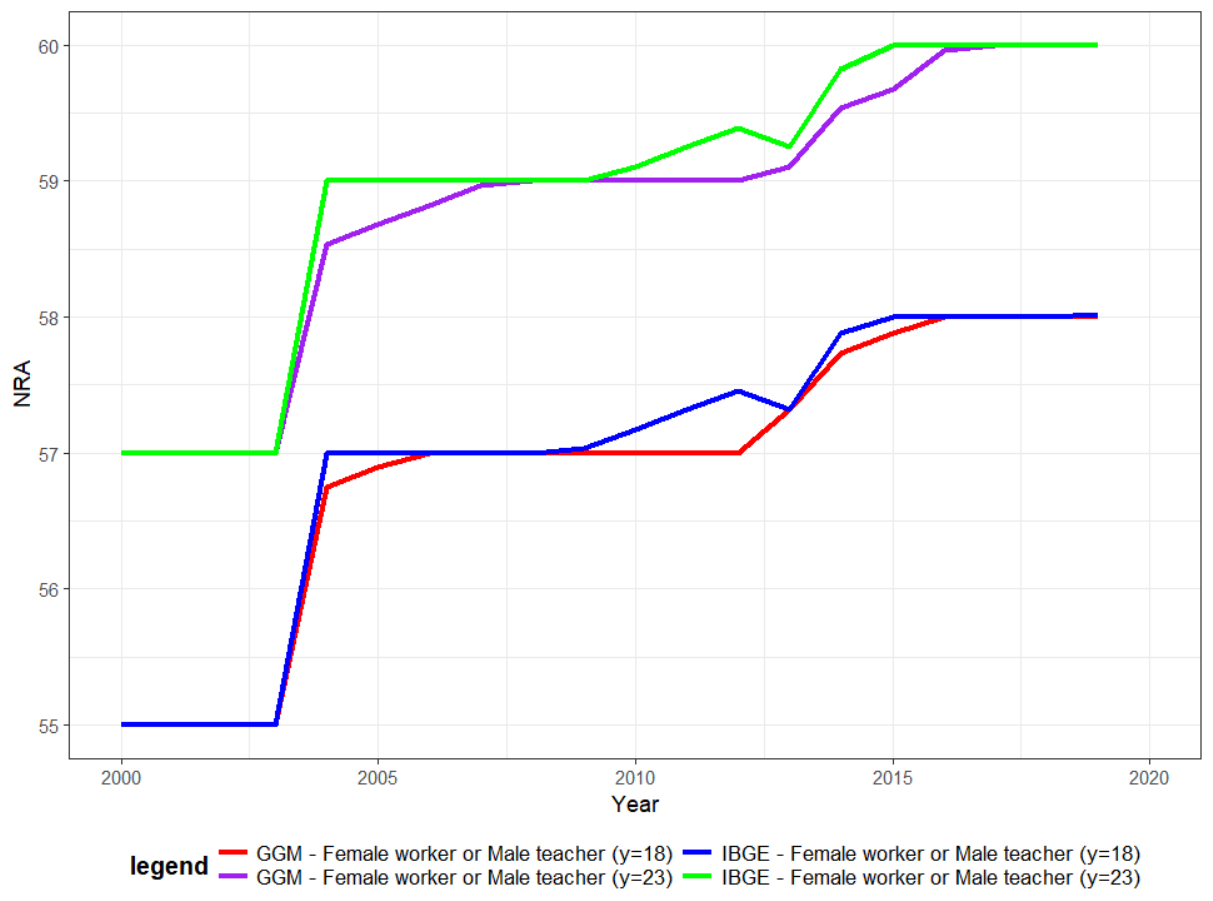}
		\caption{}\label{fig5b}
	\end{subfigure}	
	\hfill
	\begin{subfigure}{0.48\textwidth}
		\centering
		\includegraphics[width=\textwidth]{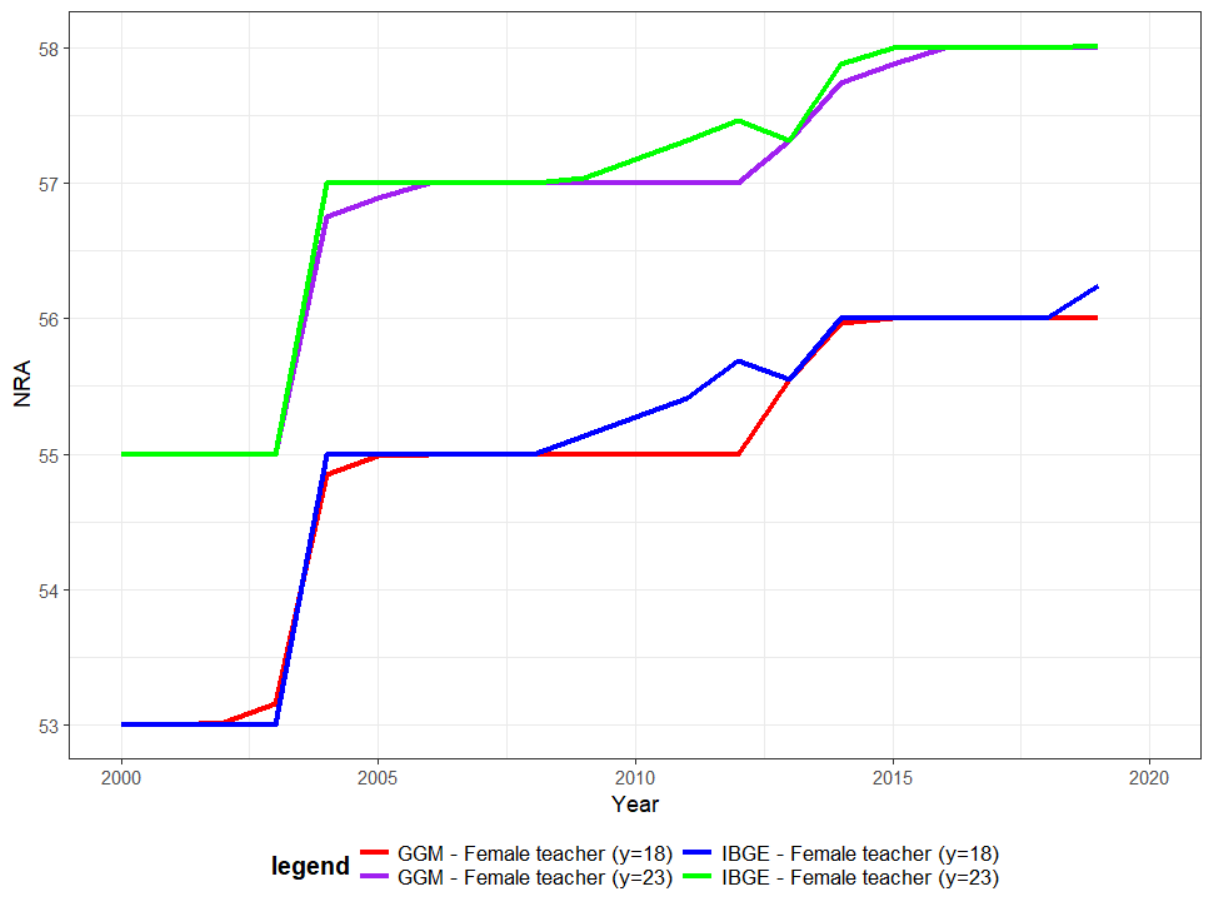}
		\caption{} \label{fig5c}
	\end{subfigure}	
	\caption{Comparison between the NRA for a male worker (a), a female worker or a male teacher (b) and a female teacher (c), from IBGE life tables and the fitted $\Gamma GM$ models, from 2000 to 2019.}
	\label{fig:NRA}
\end{figure}

The analysis of the NRA is summarized in Figure~\ref{fig:NRA}. The plot in Figure~\ref{fig5a} compares the NRAs for a male worker, Figure~\ref{fig5b} considers the case of a female worker or a male teacher, and the comparison for the case of a female teacher is shown in Figure~\ref{fig5c}. The analysis considers the results obtained from IBGE life tables and the fitted $\Gamma GM$ models, and refer to the period from 2000 to 2019.

When analyzing the NRA based on the official life tables and the SSF rule, it can be observed that, for a male worker that entered in the labor market at the age of 18 years, the NRA was around 57 years from 2000 to 2003, raising to almost 59 years from 2004 to 2009. From 2010 to 2014 the NRA gradually increased (with the exception of 2013 that used the 2011 life table) from 59.11 to 59.82 years, reaching the level of 60 years from 2015. Additionally, as mentioned in section \ref{sec:analyses}, the NRA definition assumes a whole career, i.e. $ECT=NRA-y$,  which increased from approximately 39 years in 2000 to 42 from 2015. By the $\Gamma GM$ fitted models, from 2000 to 2003, the NRA for a male worker that entered in the labor market at the age of 18 years would also be equal to 57 years. From 2004 to 2007 it gradually increased from 58.53 to 58.97, reaching 59 years in 2008 and remaining on this level until 2012. Then, it gradually increased from 59.11 in 2013 to 59.96 in 2016, reaching the level of 60 years from 2015. As shown in Figure \ref{fig5a}, the most significant difference between the NRA computed based on the IBGE tables and the counterfactual scenario was observed in 2004 (0.47 years). Moreover, since, in practice, the SSF rule uses $\lfloor x \rfloor$ when gathering the life expectancy, this impairs a more accurate calculation of the NRA. 

For a female worker or a male teacher who entered in the labor market at the age of 18 years and received a bonus of 5 years to be added to the $ECT$ when computing the SSF based on the official life tables (Figure \ref{fig5b}), the NRA was almost 55 years from 2000 to 2003 and 57 years from 2004 to 2007. From 2008 to 2014 it gradually increased from 57.03 to 57.88 (except in 2013), reaching 58 years in 2015 and remaining on this level until 2018, and had a slight increase in 2019 to 58.02. By the $\Gamma GM$ approach, the NRA was almost 55 years from 2000 to 2003 and 56.75 and 56.89 years in 2004 and 2005, respectively. In 2006 it reached 57 years remaining at this level until 2012, then it gradually increased until achieving 58 years in 2015. Moreover, the most significant difference between the NRA computed based on the IBGE tables and the counterfactual scenario occurred in 2012 (0.46 years).     

Finally, in Figure \ref{fig5c}, for a female teacher who joined the labor market at the age of 18 years and received a bonus of 10 years to be added to the ECT when computing the SSF based on the official life tables, the NRA was almost 53 years from 2000 to 2003 and 55 years from 2004 to 2008. From 2009 to 2012 it gradually increased from 55.13 to 55.69, and in 2012 it decreased to 55.55, reached 56 years in 2014, and remained at this level until 2018. In 2019 it increased again to 56.21 years. By choosing the $\Gamma GM$ estimates, the NRA was almost 53 years in 2000 and 2001 and 53.01 and 53.16 years in 2002 and 2003, respectively. In 2004 it was 54.85 years, reaching 55 years in 2006 and remaining at this level until 2012, then it gradually increased until achieving 56 years in 2015. The most considerable difference between the two scenarios was observed in 2012 (0.69 years).       

The stability of life expectancy patterns in the $\Gamma GM$ fitted models has a direct impact on the NRA, as compared to the patterns obtained from IBGE tables.\footnote{It is important to note that the IBGE life tables underwent two major methodological updates in 2002 and 2011 and that the 2002 life table was used to compute the 2004 SSF and the 2011 life table was used to compute the 2013 SSF.} Therefore, it can be noticed that the improvement of the NRA values was lower between 2003 to 2004 by the $\Gamma GM$ fitted model than by the official tables and that the NRA never reduces its value over time by the $\Gamma GM$ models. Still, it happened between 2012 and 2013 to the IBGE tables. This can be pointed out as an advantage that the $\Gamma GM$ model could have brought to the social security long-run planning. Furthermore, in all cases, when the entering age was increased from 18 to 23 years, the NRA also increased approximately by 2 years.  

It is essential to highlight that based on the 85/95 progressive rule, the NRA for a citizen that entered the labor market with 18 years would be equal to 56.5 years for a male worker (with $ECT=38.5$), 51.5 for a female worker (with $ECT=31.5$), 54 years for a male teacher (with $ECT=36$) and 49 for a female teacher (with $ECT=31$) from 2015 to 2018. Additionally, in 2019, when the rule was updated to 86/96 points, the NRA and the $ECT$ increased by 0.5 years each. Moreover, if the entering age was 23 years, both NRA and $ECT$ would increase by 2.5 years. Thus, the 85/95 progressive rule led the NRA to levels lower than or equal to it was at the beginning of the century by the SSF rule, reinforcing the fact that it was a populist policy.

\begin{figure}[h]
	\centering
	\includegraphics[scale=0.52]{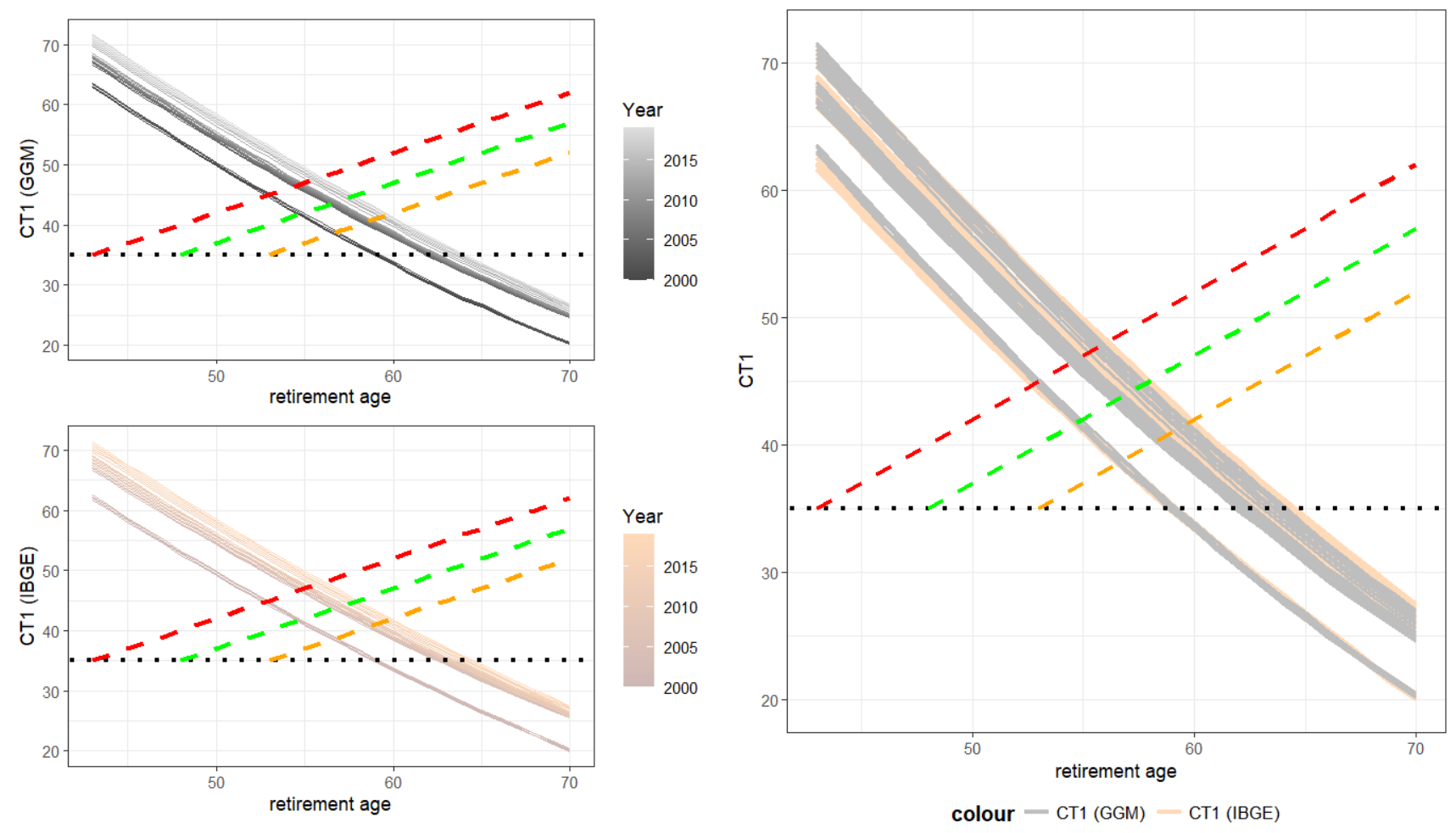}
	\caption{Contribution time needed to make the SSF equal to 1.}
	\label{fig6}
\end{figure}

Figure~\ref{fig6} shows the plots of the contribution time ($CT = CT1$) obtained by solving the Equation~\eqref{CT1} versus the integer retirement age in the interval $[43,70]$ years and for each calendar year from 2000 to 2019. The left side of Figure~\ref{fig6} shows the $\Gamma GM$ estimate (upper-left plot) and the IBGE ones (lower-right plot). On the right side, both methods are compared simultaneously. On the charts, you can see three specific lines showing the maximum contributions for various types of workers. The orange line represents male workers, the green line represents female workers and male teachers, and the red line defines female teachers. We included a horizontal black dotted line in the graph, a reference for 35 years of contribution ($CT1=35$). This line requires $CT = ECT + \beta$ to be at least 35 years. 

To obtain the positive slope lines in the plots, a constraint is solved to ensure that each retirement age has a maximum contribution time, i.e., we solved the equation $x-CT+\beta \geq 18$. It is noticeable that when the constraint is not met (i.e., values above the dashed lines), $CT1$ becomes impractical for that worker type. For example, for a female teacher that entered the labor market at 18 years and had an entire career, she could retire by the time of contribution with 43 years, having a $CT$ equal to 35 (i.e., 25 years of adequate contribution time plus 10 years of bonus). At age 44, her maximum possible $CT$ would be 36, and so on. If the $CT1$ is less than 35, paying attention to the black dotted line is important. If a worker retires at the age of contribution with $CT \geq 35$, they will have an SSF greater than 1.

From Figure~\ref{fig6}, it can be seen that, from 2000 to 2019, for any worker, it was not possible to retire with an SSF equal to 1 before the age of 53. Furthermore, from age 65, any worker retiring by the time of contribution would already have their SSF greater than or equal to 1. Therefore, we must focus on retirement ages between 53 and 64 years. Albeit Figure~\ref{fig6} are very similar at first glance, a more detailed analysis highlights essential differences in the $CT1s$, especially from 2004 to 2012 and for the highest ages.

In Table~\ref{tab:CT_1}, we can see the theoretical $CT$ that results in SSF being equal to one ($CT1$) for individuals aged 53-64. Equation~\eqref{CT1} provides more details on this. It is important to note that a female teacher at the age of 53 has a maximum $CT$ of 45 while considering the set of a male teacher or a female worker, it has a maximum $CT$ of 40, and a male worker would have a maximum $CT$ of 35. Furthermore, when the retirement age increases by one year, the maximum $CT$ increases by one year. For instance, at age 64, the maximum $CT$ values for these groups are, respectively, 56, 51, and 46. At the age of 53, from 2000 to 2003, only female teachers would have a $\textrm{SSF} \geq 1$ by the IBGE tables. However, considering the $\Gamma GM$ approach, this would only be possible in 2000 and 2001, since 2002 the $CT1$ was greater than the maximum possible $CT$ (45 years). In the period from 2000 to 2003, for the set of female workers or male teachers, it would only be possible to retire with $\textrm{SSF}=1$ at the age of 55 in both scenarios and, for a male worker, it was only possible at the age of 57. From 2004, none of the scenarios could retire with an SSF equal to 1 (by the SSF rule) with those original ages.

Between 2004 and 2012, there was a significant difference in the SSF between the two scenarios, which resulted from the differences in life expectancies. This implied in more significant differences in $CT1s$, especially for ages 59 to 64. According to IBGE life tables, a female teacher with an entire career could have retired with an SSF of 1 between 2004 and 2008. However, if the $\Gamma GM$ models were adopted, it would have been possible for a full-career female teacher to receive the full benefit until 2012. Similar results were seen for female workers and male teachers at age 57, and male workers at 59. More exact values can be found in Table~\ref{tab:CT_1}. For individuals aged between 60 and 64, retirement with an SSF of 1 was possible for any worker if their $CT$ was equal to or greater than the respective $CT1$ presented in Table~\ref{tab:CT_1}. This age group showed the most significant differences between the CT1s calculated from IBGE life tables and $\Gamma GM$ models. Between 2004 and 2012, the slightest difference between $CT1s$ was 0.7 years; however, for a person retiring at age 64 between 2005-2012, the difference could be more than 1 year. Between 2009 and 2012, a noticeable gap of 1.2 years or more between the CT1 calculated from the IBGE life tables and those calculated from the fitted $\Gamma GM$ models for a 64-year-old individual. This indicates that a worker retiring by time of contribution at the age 64, considering the official life tables, would have to work approximately 1.22 additional years than he or she would if the $\Gamma GM$ model was considered to be entitled to a SSF equal to 1.

In 2013, the official $CT1s$ were reduced compared to the previous year due to an update in the methodology of the IBGE tables. However, the $CT1s$ computed from the $\Gamma GM$ models remained stable, which highlights the reliability of these models over the IBGE life tables. Even after a methodological review, in 2013 and 2014 there were still differences in the $CT1s$ between the two models, with some cases showing a difference of more than 0.5 years. This highlights how the closing procedure adopted during this period impacted retirement metrics and may have affected workers' retirement planning. It is worth noting that the 85/95 progressive rule, implemented in mid-2015, allowed retirement with full benefits at much younger ages and lower $ECT$, as long as the minimum contribution time had been met and the retirement age and $ECT$ reached the points established by the rule.

 \begin{table}[htb!]
 	\caption{Contribution time needed to make the SSF equal to one.}
 	\resizebox{\textwidth}{!}{%
 		\centering
 		\begin{tabular}{lcccccccccccc}
 			\hline
 			\multicolumn{13}{c}{IBGE}\\
 			\hline
 			Year/Age& 53& 54& 55& 56& 57& 58& 59& 60& 61& 62& 63& 64\\
 			\hline
 			2000& 44.13& 42.63& 40.97& 39.32& 37.85& 36.22& 34.77& 33.33& 31.90& 30.48& 29.07& 27.66\\
 			2001& 44.48& 42.81& 41.15& 39.50& 38.03& 36.40& 34.95& 33.51& 32.08& 30.66& 29.25& 27.84\\
 			2002& 44.66& 42.99& 41.33& 39.85& 38.21& 36.75& 35.13& 33.69& 32.26& 30.84& 29.42& 28.02\\
 			2003& 44.84& 43.17& 41.51& 40.03& 38.39& 36.93& 35.31& 33.87& 32.44& 31.01& 29.60& 28.20\\
 			2004& 49.27& 47.60& 46.11& 44.46& 42.99& 41.36& 39.91& 38.46& 37.03& 35.61& 34.37& 32.96\\
 			2005& 49.44& 47.95& 46.29& 44.63& 43.17& 41.71& 40.08& 38.64& 37.21& 35.78& 34.37& 33.13\\
 			2006& 49.79& 48.12& 46.46& 44.99& 43.34& 41.88& 40.26& 38.81& 37.38& 35.96& 34.54& 33.31\\
 			2007& 49.97& 48.30& 46.64& 45.16& 43.52& 42.06& 40.43& 38.99& 37.56& 36.13& 34.72& 33.48\\
 			2008& 50.14& 48.47& 46.99& 45.34& 43.69& 42.23& 40.78& 39.17& 37.73& 36.31& 34.89& 33.66\\
 			2009& 50.32& 48.83& 47.16& 45.51& 44.04& 42.41& 40.96& 39.52& 37.91& 36.48& 35.07& 33.83\\
 			2010& 50.67& 49.00& 47.34& 45.69& 44.22& 42.58& 41.13& 39.69& 38.08& 36.66& 35.24& 34.01\\
 			2011& 50.84& 49.17& 47.51& 46.04& 44.39& 42.76& 41.31& 39.86& 38.26& 36.83& 35.41& 34.18\\
 			2012& 51.02& 49.35& 47.86& 46.21& 44.57& 43.11& 41.48& 40.04& 38.61& 37.01& 35.59& 34.35\\
 			2013& 51.02& 49.35& 47.69& 46.04& 44.39& 42.93& 41.31& 39.69& 38.26& 36.66& 35.24& 33.83\\
 			2014& 51.72& 50.05& 48.39& 46.73& 45.09& 43.63& 42.01& 40.39& 38.95& 37.35& 35.94& 34.53\\
 			2015& 52.07& 50.40& 48.74& 47.08& 45.44& 43.81& 42.35& 40.74& 39.13& 37.70& 36.28& 34.70\\
 			2016& 52.41& 50.74& 49.08& 47.43& 45.79& 44.15& 42.53& 41.08& 39.48& 37.88& 36.46& 35.05\\
 			2017& 52.76& 51.09& 49.26& 47.61& 45.96& 44.50& 42.88& 41.26& 39.65& 38.22& 36.81& 35.22\\
 			2018& 52.94& 51.27& 49.61& 47.95& 46.31& 44.68& 43.05& 41.61& 40.00& 38.40& 36.98& 35.57\\
 			2019& 53.28& 51.61& 49.95& 48.30& 46.66& 45.02& 43.40& 41.78& 40.17& 38.74& 37.15& 35.74\\
 			\hline
 			\multicolumn{13}{c}{$\Gamma GM$}\\
 			\hline
 			2000& 44.48& 42.81& 41.15& 39.50& 37.85& 36.40& 34.77& 33.33& 31.72& 30.30& 28.89& 27.49\\
 			2001& 44.84& 42.99& 41.33& 39.68& 38.03& 36.58& 34.95& 33.51& 31.90& 30.48& 29.07& 27.66\\
 			2002& 45.02& 43.35& 41.51& 39.85& 38.39& 36.75& 35.13& 33.69& 32.08& 30.66& 29.25& 27.84\\
 			2003& 45.20& 43.53& 41.86& 40.21& 38.57& 36.93& 35.31& 33.87& 32.44& 30.84& 29.42& 28.02\\
 			2004& 48.74& 47.07& 45.41& 43.93& 42.29& 40.65& 39.20& 37.76& 36.33& 34.73& 33.49& 32.08\\
 			2005& 48.92& 47.25& 45.58& 44.11& 42.46& 40.83& 39.38& 37.94& 36.33& 34.90& 33.49& 32.08\\
 			2006& 49.09& 47.42& 45.76& 44.28& 42.64& 41.00& 39.55& 37.94& 36.50& 35.08& 33.66& 32.26\\
 			2007& 49.27& 47.60& 45.94& 44.28& 42.82& 41.18& 39.73& 38.11& 36.68& 35.26& 33.84& 32.43\\
 			2008& 49.44& 47.77& 46.11& 44.46& 42.99& 41.36& 39.73& 38.29& 36.86& 35.43& 33.84& 32.61\\
 			2009& 49.62& 47.95& 46.29& 44.63& 43.17& 41.53& 39.91& 38.46& 37.03& 35.43& 34.02& 32.61\\
 			2010& 49.79& 48.12& 46.46& 44.81& 43.34& 41.71& 40.08& 38.64& 37.21& 35.61& 34.19& 32.78\\
 			2011& 50.14& 48.30& 46.64& 44.99& 43.52& 41.88& 40.26& 38.81& 37.21& 35.78& 34.37& 32.96\\
 			2012& 50.32& 48.47& 46.81& 45.16& 43.69& 42.06& 40.43& 38.99& 37.38& 35.96& 34.54& 33.13\\
 			2013& 51.19& 49.52& 47.69& 46.04& 44.39& 42.76& 41.13& 39.69& 38.08& 36.48& 35.07& 33.66\\
 			2014& 51.72& 49.87& 48.21& 46.56& 44.92& 43.28& 41.66& 40.04& 38.43& 37.01& 35.41& 34.01\\
 			2015& 51.89& 50.22& 48.56& 46.73& 45.09& 43.46& 41.83& 40.21& 38.78& 37.18& 35.76& 34.35\\
 			2016& 52.24& 50.40& 48.74& 47.08& 45.44& 43.81& 42.18& 40.56& 38.95& 37.53& 35.94& 34.53\\
 			2017& 52.59& 50.74& 49.08& 47.26& 45.61& 43.98& 42.35& 40.74& 39.30& 37.70& 36.28& 34.70\\
 			2018& 52.76& 51.09& 49.26& 47.61& 45.96& 44.33& 42.70& 41.08& 39.48& 37.88& 36.46& 34.88\\
 			2019& 53.11& 51.27& 49.61& 47.78& 46.14& 44.50& 42.88& 41.26& 39.65& 38.22& 36.63& 35.22\\
 			\hline
 			\label{tab:CT_1} 
 	\end{tabular}}
 \end{table}


\section{Discussion}\label{sec:Discussion}

As per \cite{Caetano2016}, the SSF was introduced in 1999 as an imperfect substitute for the failed attempt to establish a minimal retirement age rule in Brazil. However, it has faced criticism for being complexity, lacking theoretical actuarial background \citep{Beltrao2014}, causing financial losses for retirees, and failing to balance the system's accounts \citep{Lima2012}. Because of this, the SSF rule suffered continuous pressure of further reforms until it was replaced by a statutory retirement age rule in 2019. Nonetheless, \cite{Queiroz2021} highlights that the new reform does not implement any AAM, which means that it is unlikely to have a long-term effect and, consequently, new reforms will be necessary in the future. Therefore, understanding some limitations of the SSF can help in the planning of a new AAM and prevent mistakes from being repeated, especially because, as stated, past experiences would most likely be remembered in new political debates on pension reform.

Due to the focus of the paper, the first point that should be highlighted about the use of life tables to compute the SSF is the fact that, even though retirement age could (and probably is) a non-integer age, IBGE life tables only provide life expectancies for integer ages, and the retirement age needed to be rounded down to the nearest integer age so that the life expectancy could be properly obtained. However, since life expectancy is typically a monotonically decreasing function of age \citep{Canudas2011}, the reduction of the retirement age to the nearest integer age could cause a reduction on the SSF and, potentially, on retiree's income. In this sense, since life expectancy can be computed for any age directly from the parameters of the $\Gamma GM$ fitted model \citep{Castellares2020}, then the SSF could be calculated more accurately, which would benefit retirees. Although some might suggest using a fractional age assumption such as the UDD \citep{Dickson2019} as a workaround for this limitation, a parametric mortality law has the advantage of not requiring any additional assumption, maintaining the consistency of the mortality pattern. Furthermore, the possibility of using non-integer ages would also benefit the calculation of metrics like the NRA, which would also be important from the public planner's point of view.

Another issue concerning the use of life expectancy in AAM is related to its updates. \cite{Delgado2006} and \cite{Wolff2011} advert that a particular metric (such as life expectancy) published in a given calendar year should be seen as an inexact measure, that could be subject to revision due to data and methodological updates. This issue raises two main questions: (i) what should be done when revised tables are released?, i.e, should retirees also have their pensions revised (whether for higher or lower values)?, and (ii) what strategies could be adopted to mitigate sudden variations on the adjustment mechanism? The first question is essential since data updates can potentially cause long legal disputes and create uncertainty for both the government and citizens. As a result, it would be beneficial for legislation to anticipate and outline how the mechanism should respond in the event of revised data disclosure, particularly if such data could benefit retirees. The second question is also crucial because sudden fluctuations may disrupt long-term planning and diminish popular and political support to the AAM. This could ultimately result in abandoning the AAM and implementing discretionary or populist measures, as those we have seen in Brazil since the SSF rule was implemented.
 
According to \cite{Tavernier2021}, using a moving average of life expectancy over several years could help to create a stable and predictable mechanism. Moreover, this approach could also help to reduce fluctuations caused by environmental factors like the COVID-19 pandemic \citep{Castro2021}. This strategy combined with a law that outlines how to react to revised data disclosure, could also protect statistical agencies from political pressure, allowing them to always adopt the best data and methods available. In addition, parametric mortality laws, such as the $\Gamma GM$  model, are less sensible to the censoring age \citep{Missov2016}, which could be seen as an additional strategy to maintain the mechanism more stable. However, it is important to note that using different distributions or estimation strategies \citep{Castellares2022} or even changing the model used to fit the data (or the adoption of mixture models) could impact life expectancy calculation \citep{Cabral2022}. This means that parametric models are not immune to data and methodological updates.

Previous criticisms to the SSF also pointed out that the life expectancies used in its calculation should be those that best describe the survival pattern of the system's beneficiaries, and not those describing the whole Brazilian population \citep{Cechin2007, Wilbert2013}. \cite{Ribeiro2008} complement this argument indicating that the SSF overlooks the mortality heterogeneity of Brazilian population, potentially exacerbating social inequalities and causing regressive effects to the system. Future research should investigate how different social groups are impacted when AAM uses aggregated indicators, particularly if the mortality gap between lower and higher-income groups continue to widen \citep{Sanchez2020}. This is highlighted in recent studies by \cite{Haan2020} and \cite{Alvarez2021}. In this respect, administrative micro-data from INSS could be employed to estimate old-age mortality for different groups, as done by \cite{Gonzaga2022}. The authors stated that, in 2015, life  expectancies at 65 and 75 years for the INSS beneficiaries were higher than those provided by the IBGE for the general population, especially for women, and they also reinforced the importance of new studies on mortality differentials between socioeconomic groups and the need of continuous efforts to improve mortality data quality, particularly at old ages. 

\cite{Queiroz2017, Queiroz2020} highlight the importance of completeness of death registration to the proper construction of life tables, and observed that although the quality of mortality data has improved, IBGE completeness of death estimates showed the lowest levels for the Northern of the country compared to other estimates \citep{Queiroz2020}. \cite{Gomes2009} and \cite{Nepomuceno2020} also shared their concerns about the low-quality data about the size of old-age population in Brazil, arguing that this problem may cause severe implications in the estimation of adult mortality. Therefore, those are important topics to be addressed by IBGE and other research centers in Brazil, especially because life expectancy is a critical indicator to social security planning.  
 
One of the most contemporary issues in the debate about adopting of life expectancies in AAM is the difference between cohort and period life expectancies. Period life expectancy at a given age $x$ is computed from a period life table which assumes that age-specific death rates (for a given calendar year) will remain constant over time, and therefore it does not reflect the average number of years that a person aged $x$ will be expected to live in the real population \citep{Guerra2013}. Cohort life expectancies, on its turn, are typically computed from cohort tables constructed based on past mortality experience and on future age-specific death rate projections that try to account for longevity improvements, seeking to express the average length of life of a real cohort of citizens born at the same year \citep{Goldstein2006, Luy2020, Modig2020}. Studies such as \cite{Ayuso2021, Ayuso2021b} and \cite{Bravo2021} warn about adopting cohort life expectancies in AAM due to the longevity improvements observed in the past decades. According to the authors, the adoption of period life tables by the pension system causes a systematic underestimation of life expectancy, which results in subsidies from future to current generations. However, the authors also advert that cohort life tables demand more assumptions than period ones to be constructed, particularly about future mortality trends, which can be an additional difficulty for statistical agencies and in the political debate. In this sense, future studies investigating the impact that cohort life expectancy could have had on the SSF would undoubtedly shed light on the advantages and disadvantages of adopting cohort tables on future AAM by Brazilian legislation.

It is essential to be aware that social security planning is not a simple task since social security systems typically have multiple goals such as social protection, consumption smoothing, and income distribution, among others \citep{Barr2010}. It involves long-run analyses, which may be affected by forecast assumptions \citep{Silva2017}, and politicians may have short-run goals or be influenced by popular pressure \citep{Vidal2009}. Imperfect information and financial illiteracy may lead workers to make wrong and costly retirement decisions \citep{Barr2006}, affecting both their wealth and the system's balancing, which forces legislators to choose between a more flexible or a more stringent regulation \citep{VanVuuren2014}. Demographic dynamics imposes fiscal challenges \citep{Freitas2019}, and the system's design has implications on adult and old-age participation in the labor market \citep{Queiroz2021, Souza2019} and on savings, and, consequently, on national economic performance \citep{Barr2010}. That is why it should be emphasized that AAM that links the pension benefit to the life expectancy at the retirement age, as is the case of the SSF, traditionally needs to be combined with additional adjustment mechanisms, such as minimal retirement age AAM, or even automatic balancing mechanisms that monitor the retiree-workers ratio \citep{Tavernier2021}, for example, to try to accommodate all these issues.  

Finally, it is also crucial to highlight that IBGE is a well-known and respected statistical agency due to its autonomy and quality of its human capital \citep{Nepomuceno2020}, as well as for its constant search to improve its practices based on rigorous research, and in dialogue with international agencies. The IBGE \citep{IBGE2004} already recognizes that different approaches to constructing life tables, however similar they may seem, can generate different results. Therefore, with this study, we hope to stimulate the debate about the adoption of life expectancy in AAM in Brazil and to help public planners to improve their practices, having the consciousness that even apparently harmless hypotheses can have sizable impacts on the results of public actions. 

	
\section{Conclusion}\label{sec:Conclusion}

The use of life-table metrics (particularly life expectancies) computed from national life tables in AAM to help social security systems properly respond to longevity improvements is one of many indisputable proofs of the importance that life tables had for public planning. In this paper, we discuss how life-table closing procedures affected an AAM that links retirement benefits to life expectancy in the Brazilian social security context. Our results indicate that, from 2004 to 2012, the adoption of official life tables to compute life expectancy may have caused sizable negative impacts on retirees' income, especially for those who delayed their retirement. Moreover, the life expectancies provided by the $\Gamma GM$ fitted models had more stable paths over time than those computed from official tables, which could have helped long-term planning. Finally, we understand that this study also contributes to warning public planners about the impacts that seemingly innocuous hypotheses could have on the results of public actions.

\section*{Funding}
Filipe Costa de Souza and Wilton Bernardino thank Federal University of Pernambuco through the notice 09/2021.

Silvio C. Patricio gratefully acknowledges the support from AXA Research Fund through funding the ”AXA Chair in Longevity Research.” 

Wilton Bernardino thanks the support of the National Council for Scientific and Technological Development (CNPq) through the Grants 200615806, 210616772, 210616773, 220618257, and 220618286. 

The funders had no role in study design, data collection and analysis, decision to publish, or preparation of the manuscript.

\section*{Conflict of interest}
The authors declare that they have no conflict of interest.

\bibliographystyle{apalike}
\bibliography{reference}

\begin{appendices}\label{sec:appendix}

	\begin{landscape}
		
		\begin{table}[h]
			\caption{Fragment of the 2012 Social Security Factor table built from the 2010 IBGE life table}
			\resizebox{\linewidth}{!}{%
				\centering
				\begin{tabular}{lccccccccccccccccccc}
					\hline
					CT/Age & 43& 44& 45& 46& 47& 48& 49& 50& 51& 52& 53& 54& 55& 56& 57& 58& 59& 60\\
					\hline
					35& 0.476& 0.491& 0.506& 0.524& 0.540& 0.560& 0.578& 0.598& 0.618& 0.643& 0.666& 0.691& 0.714& 0.742& 0.772& 0.800& 0.834& 0.866\\
					36& & 0.506& 0.522& 0.540& 0.557& 0.577& 0.596& 0.616& 0.637& 0.662& 0.686& 0.712& 0.736& 0.765& 0.795& 0.824& 0.859& 0.893\\
					37& & & 0.537& 0.556& 0.573& 0.594& 0.614& 0.634& 0.656& 0.682& 0.707& 0.733& 0.758& 0.787& 0.819& 0.849& 0.885& 0.919\\
					38& & & & 0.572& 0.590& 0.611& 0.631& 0.653& 0.675& 0.702& 0.727& 0.754& 0.780& 0.810& 0.842& 0.873& 0.910& 0.946\\
					39& & & & & 0.607& 0.628& 0.649& 0.671& 0.694& 0.721& 0.748& 0.775& 0.802& 0.833& 0.866& 0.898& 0.936& 0.972\\
					40& & & & & & 0.646& 0.667& 0.690& 0.713& 0.741& 0.768& 0.797& 0.824& 0.856& 0.890& 0.923& 0.962& 0.999\\
					41& & & & & & & 0.685& 0.708& 0.733& 0.761& 0.789& 0.818& 0.846& 0.879& 0.914& 0.947& 0.988& 1.026\\
					42& & & & & & & & 0.727& 0.752& 0.781& 0.810& 0.840& 0.868& 0.902& 0.938& 0.972& 1.013& 1.053\\
					43& & & & & & & & & 0.771& 0.801& 0.830& 0.861& 0.890& 0.925& 0.962& 0.997& 1.039& 1.080\\
					44& & & & & & & & & & 0.822& 0.851& 0.883& 0.913& 0.948& 0.986& 1.022& 1.066& 1.107\\
					45& & & & & & & & & & & 0.872& 0.905& 0.935& 0.972& 1.010& 1.047& 1.092& 1.134\\
					46& & & & & & & & & & & & 0.926& 0.958& 0.995& 1.035& 1.073& 1.118& 1.161\\
					47& & & & & & & & & & & & & 0.980& 1.019& 1.059& 1.098& 1.144& 1.189\\
					48& & & & & & & & & & & & & & 1.042& 1.084& 1.123& 1.171& 1.216\\
					49& & & & & & & & & & & & & & & 1.108& 1.149& 1.197& 1.244\\
					50& & & & & & & & & & & & & & & & 1.174& 1.224& 1.271\\
					51& & & & & & & & & & & & & & & & & 1.251& 1.299\\
					52& & & & & & & & & & & & & & & & & & 1.327\\
					\hline
					\label{tab:SSF_IBGE}
			\end{tabular}}
		\end{table}
	\end{landscape}
	
	\begin{landscape}
		
		\begin{table}[h]
			\caption{Fragment of the 2012 Social Security Factor table built from the 2010 $\Gamma GM$ fitted model}
			\resizebox{\linewidth}{!}{%
				\centering
				\begin{tabular}{lccccccccccccccccccc}
					\hline
					CT/Age& 43& 44& 45& 46& 47& 48& 49& 50& 51& 52& 53& 54& 55& 56& 57& 58& 59& 60\\
					\hline
					35& 0.478& 0.494& 0.511& 0.529& 0.547& 0.565& 0.586& 0.606& 0.629& 0.652& 0.676& 0.704& 0.731& 0.761& 0.788& 0.822& 0.857& 0.891\\
					36& & 0.509& 0.527& 0.545& 0.564& 0.582& 0.604& 0.624& 0.649& 0.672& 0.697& 0.726& 0.754& 0.784& 0.812& 0.847& 0.883& 0.918\\
					37& & & 0.542& 0.561& 0.581& 0.600& 0.622& 0.643& 0.668& 0.692& 0.717& 0.747& 0.776& 0.807& 0.837& 0.872& 0.909& 0.946\\
					38& & & & 0.577& 0.598& 0.617& 0.640& 0.662& 0.687& 0.712& 0.738& 0.769& 0.799& 0.830& 0.861& 0.897& 0.936& 0.973\\
					39& & & & & 0.615& 0.635& 0.658& 0.680& 0.707& 0.732& 0.759& 0.791& 0.821& 0.854& 0.885& 0.922& 0.962& 1.000\\
					40& & & & & & 0.652& 0.676& 0.699& 0.726& 0.752& 0.780& 0.812& 0.844& 0.877& 0.909& 0.948& 0.989& 1.028\\
					41& & & & & & & 0.694& 0.718& 0.746& 0.772& 0.801& 0.834& 0.867& 0.901& 0.934& 0.973& 1.015& 1.055\\
					42& & & & & & & & 0.737& 0.765& 0.793& 0.822& 0.856& 0.889& 0.925& 0.958& 0.999& 1.042& 1.083\\
					43& & & & & & & & & 0.785& 0.813& 0.843& 0.878& 0.912& 0.948& 0.983& 1.024& 1.068& 1.111\\
					44& & & & & & & & & & 0.834& 0.864& 0.900& 0.935& 0.972& 1.008& 1.050& 1.095& 1.139\\
					45& & & & & & & & & & & 0.886& 0.922& 0.958& 0.996& 1.032& 1.076& 1.122& 1.167\\
					46& & & & & & & & & & & & 0.945& 0.981& 1.020& 1.057& 1.102& 1.149& 1.195\\
					47& & & & & & & & & & & & & 1.004& 1.044& 1.082& 1.128& 1.176& 1.223\\
					48& & & & & & & & & & & & & & 1.068& 1.107& 1.154& 1.203& 1.251\\
					49& & & & & & & & & & & & & & & 1.132& 1.180& 1.231& 1.279\\
					50& & & & & & & & & & & & & & & & 1.206& 1.258& 1.308\\
					51& & & & & & & & & & & & & & & & & 1.285& 1.336\\
					52& & & & & & & & & & & & & & & & & & 1.365\\
					\hline
					\label{tab:SSF_GGM}
			\end{tabular}}
		\end{table}
	\end{landscape}
	
	\begin{landscape}
		
		\begin{table}[h]
			\caption{Fragment of the 2015 SSF table built from the 2013 IBGE life table, for a female worker, based on the combination of the SSF rule and the 85/95 progressive rule}
			\resizebox{\linewidth}{!}{%
				\centering
				\begin{tabular}{lcccccccccccccccccc}
					\hline
					ECT/Age& 48& 49& 50& 51& 52& 53& 54& 55& 56& 57& 58& 59& 60& 61& 62& 63& 64& 65\\
					\hline
					30& 0.547& 0.567& 0.586& 0.606& 0.629& 0.651& 0.675& 1.000& 1.000& 1.000& 1.000& 1.000& 1.000& 1.000& 1.000& 1.000& 1.009& 1.054\\
					31& & 0.584& 0.604& 0.624& 0.648& 0.671& 1.000& 1.000& 1.000& 1.000& 1.000& 1.000& 1.000& 1.000& 1.000& 1.000& 1.040& 1.086\\
					32& & & 0.621& 0.643& 0.667& 1.000& 1.000& 1.000& 1.000& 1.000& 1.000& 1.000& 1.000& 1.000& 1.000& 1.021& 1.071& 1.118\\
					33& & & & 0.661& 1.000& 1.000& 1.000& 1.000& 1.000& 1.000& 1.000& 1.000& 1.000& 1.000& 1.008& 1.050& 1.101& 1.151\\
					34& & & & & 1.000& 1.000& 1.000& 1.000& 1.000& 1.000& 1.000& 1.000& 1.000& 1.000& 1.037& 1.080& 1.132& 1.183\\
					35& & & & & & 1.000& 1.000& 1.000& 1.000& 1.000& 1.000& 1.000& 1.000& 1.024& 1.065& 1.110& 1.163& 1.215\\
					36& & & & & & & 1.000& 1.000& 1.000& 1.000& 1.000& 1.000& 1.007& 1.051& 1.094& 1.139& 1.195& 1.248\\
					37& & & & & & & & 1.000& 1.000& 1.000& 1.000& 1.000& 1.033& 1.079& 1.123& 1.169& 1.226& 1.281\\
					38& & & & & & & & & 1.000& 1.000& 1.000& 1.016& 1.060& 1.107& 1.151& 1.199& 1.257& 1.313\\
					39& & & & & & & & & & 1.000& 1.005& 1.042& 1.086& 1.134& 1.180& 1.229& 1.289& 1.346\\
					40& & & & & & & & & & & 1.029& 1.068& 1.113& 1.162& 1.209& 1.259& 1.320& 1.379\\
					41& & & & & & & & & & & & 1.093& 1.140& 1.190& 1.238& 1.290& 1.352& 1.412\\
					42& & & & & & & & & & & & & 1.167& 1.218& 1.267& 1.320& 1.384& 1.445\\
					43& & & & & & & & & & & & & & 1.246& 1.297& 1.350& 1.416& 1.479\\
					44& & & & & & & & & & & & & & & 1.326& 1.381& 1.448& 1.512\\
					45& & & & & & & & & & & & & & & & 1.412& 1.480& 1.546\\
					46& & & & & & & & & & & & & & & & & 1.512& 1.579\\
					47& & & & & & & & & & & & & & & & & & 1.613\\
					\hline
			\end{tabular}}
			Note: The number one that appears in some cells means that the condition to retire by the 85/95 progressive has been met, and that $\textrm{SSF}_{x, CT}$ computed based on Equation~\eqref{factor} is lower than or equal to 1, i.e, the worker would be in a better condition if she retired by the 85/95 progressive rule than by the SSF rule. 
			\label{tab:SSF_IBGE_85}
		\end{table}
	\end{landscape}

	\begin{landscape}
		
		\begin{table}[h]
			\caption{Fragment of the 2015 SSF table built from the 2013 $\Gamma$GM fitted, for a female worker, based on the combination of the SSF rule and the 85/95 progressive rule}
			\resizebox{\linewidth}{!}{%
				\centering
				\begin{tabular}{lcccccccccccccccccc}
					\hline
					ECT/Age& 48& 49& 50& 51& 52& 53& 54& 55& 56& 57& 58& 59& 60& 61& 62& 63& 64& 65\\
					\hline
					30& 0.547& 0.565& 0.586& 0.608& 0.629& 0.654& 0.678& 1.000& 1.000& 1.000& 1.000& 1.000& 1.000& 1.000& 1.000& 1.000& 1.020& 1.072\\
					31& & 0.582& 0.604& 0.626& 0.648& 0.674& 1.000& 1.000& 1.000& 1.000& 1.000& 1.000& 1.000& 1.000& 1.000& 1.007& 1.051& 1.104\\
					32& & & 0.621& 0.645& 0.667& 1.000& 1.000& 1.000& 1.000& 1.000& 1.000& 1.000& 1.000& 1.000& 1.000& 1.037& 1.082& 1.137\\
					33& & & & 0.664& 1.000& 1.000& 1.000& 1.000& 1.000& 1.000& 1.000& 1.000& 1.000& 1.000& 1.024& 1.067& 1.113& 1.170\\
					34& & & & & 1.000& 1.000& 1.000& 1.000& 1.000& 1.000& 1.000& 1.000& 1.000& 1.006& 1.052& 1.097& 1.145& 1.203\\
					35& & & & & & 1.000& 1.000& 1.000& 1.000& 1.000& 1.000& 1.000& 1.000& 1.034& 1.081& 1.127& 1.176& 1.236\\
					36& & & & & & & 1.000& 1.000& 1.000& 1.000& 1.000& 1.000& 1.021& 1.061& 1.110& 1.157& 1.208& 1.269\\
					37& & & & & & & & 1.000& 1.000& 1.000& 1.000& 1.004& 1.048& 1.089& 1.139& 1.187& 1.239& 1.302\\
					38& & & & & & & & & 1.000& 1.000& 1.000& 1.030& 1.075& 1.117& 1.169& 1.218& 1.271& 1.335\\
					39& & & & & & & & & & 1.000& 1.013& 1.056& 1.102& 1.145& 1.198& 1.248& 1.303& 1.369\\
					40& & & & & & & & & & & 1.038& 1.082& 1.129& 1.173& 1.227& 1.279& 1.335& 1.402\\
					41& & & & & & & & & & & & 1.108& 1.156& 1.202& 1.257& 1.310& 1.367& 1.436\\
					42& & & & & & & & & & & & & 1.183& 1.230& 1.286& 1.341& 1.399& 1.470\\
					43& & & & & & & & & & & & & & 1.258& 1.316& 1.371& 1.431& 1.504\\
					44& & & & & & & & & & & & & & & 1.346& 1.402& 1.463& 1.538\\
					45& & & & & & & & & & & & & & & & 1.434& 1.496& 1.572\\
					46& & & & & & & & & & & & & & & & & 1.528& 1.606\\
					47& & & & & & & & & & & & & & & & & & 1.640\\
					\hline
					\label{tab:SSF_GGM_85}
			\end{tabular}}
		\end{table}
	\end{landscape}	
	
	\begin{landscape}
		
		\begin{table}[h]
			\caption{Fragment of the relative discrepancy (\%) between the 2015 SSF built from the 2013 IBGE life table and the one built from the 2013 $\Gamma$GM fitted model, for a female worker, based on the combination of the SSF rule and the 85/95 progressive rule}
			\resizebox{\linewidth}{!}{%
				\centering
				\begin{tabular}{lcccccccccccccccccc}
					\hline
					ECT/Age& 48& 49& 50& 51& 52& 53& 54& 55& 56& 57& 58& 59& 60& 61& 62& 63& 64& 65\\
					\hline
					30& 0.000& -0.326& 0.000& 0.346& 0.000& 0.368& 0.379& 0.000& 0.000& 0.000& 0.000& 0.000& 0.000& 0.000& 0.000& 0.000& 1.075& 1.685\\
					31& & -0.326& 0.000& 0.346& 0.000& 0.368& 0.000& 0.000& 0.000& 0.000& 0.000& 0.000& 0.000& 0.000& 0.000& 0.706& 1.075& 1.685\\
					32& & & 0.000& 0.346& 0.000& 0.000& 0.000& 0.000& 0.000& 0.000& 0.000& 0.000& 0.000& 0.000& 0.000& 1.554& 1.075& 1.685\\
					33& & & & 0.346& 0.000& 0.000& 0.000& 0.000& 0.000& 0.000& 0.000& 0.000& 0.000& 0.000& 1.500& 1.554& 1.075& 1.685\\
					34& & & & & 0.000& 0.000& 0.000& 0.000& 0.000& 0.000& 0.000& 0.000& 0.000& 0.609& 1.500& 1.554& 1.075& 1.685\\
					35& & & & & & 0.000& 0.000& 0.000& 0.000& 0.000& 0.000& 0.000& 0.000& 0.962& 1.500& 1.554& 1.075& 1.685\\
					36& & & & & & & 0.000& 0.000& 0.000& 0.000& 0.000& 0.000& 1.395& 0.962& 1.500& 1.554& 1.075& 1.685\\
					37& & & & & & & & 0.000& 0.000& 0.000& 0.000& 0.435& 1.395& 0.962& 1.500& 1.554& 1.075& 1.685\\
					38& & & & & & & & & 0.000& 0.000& 0.000& 1.345& 1.395& 0.962& 1.500& 1.554& 1.075& 1.685\\
					39& & & & & & & & & & 0.000& 0.866& 1.345& 1.395& 0.962& 1.500& 1.554& 1.075& 1.685\\
					40& & & & & & & & & & & 0.866& 1.345& 1.395& 0.962& 1.500& 1.554& 1.075& 1.685\\
					41& & & & & & & & & & & & 1.345& 1.395& 0.962& 1.500& 1.554& 1.075& 1.685\\
					42& & & & & & & & & & & & & 1.395& 0.962& 1.500& 1.554& 1.075& 1.685\\
					43& & & & & & & & & & & & & & 0.962& 1.500& 1.554& 1.075& 1.685\\
					44& & & & & & & & & & & & & & & 1.500& 1.554& 1.075& 1.685\\
					45& & & & & & & & & & & & & & & & 1.554& 1.075& 1.685\\
					46& & & & & & & & & & & & & & & & & 1.075& 1.685\\
					47& & & & & & & & & & & & & & & & & & 1.685\\
					\hline
					\label{tab:relative difference}
			\end{tabular}}
			
		\end{table}
	\end{landscape}	
	
\end{appendices}

\end{document}